%% file: main.tex
\documentclass[manuscript, screen]{acmart}
\AtBeginDocument{%
  }

\setcopyright{acmlicensed}
\copyrightyear{2018}
\acmYear{2018}
\acmDOI{XXXX.XXX}
\acmConference{Conference}{June 03--05,
  2018}{Woodstock, NY}
\acmISBN{978-1-4503-XXXX-X/2018/06}

\input{./meta/commands.tex}

\begin{document}


\title{Inflak: A Unified Orchestration Architecture for Human-Agent Interaction}



\author{Lin Gao}
\authornote{Work done during internship at Microsoft Research Asia.}
\email{lingao23@m.fudan.edu.cn}
\orcid{0009-0004-1613-1774}
\affiliation{%
  \institution{Fudan University}
  \state{Shanghai}
  \country{China}
}

\author{Li Chen}
\email{lichen3@microsoft.com}
\affiliation{%
  \institution{Microsoft}
  \state{Washington}
  \country{USA}
}

\author{Jianxun Lian}
\email{jialia@microsoft.com}
\orcid{0000-0003-3108-5601}
\affiliation{%
  \institution{Microsoft Research Asia}
  \state{Beijing}
  \country{China}
}

\author{Ji Li}
\email{jili5@microsoft.com}
\affiliation{%
  \institution{Microsoft}
  \state{California}
  \country{USA}
}

\author{Haotian Li}
\email{haotian.li@microsoft.com}
\orcid{0000-0001-9547-3449}
\affiliation{%
  \institution{Microsoft Research Asia}
  \state{Beijing}
  \country{China}
}



\input{sections/0-abstract}




\begin{teaserfigure}
   \centering 
  \includegraphics[width=\columnwidth]{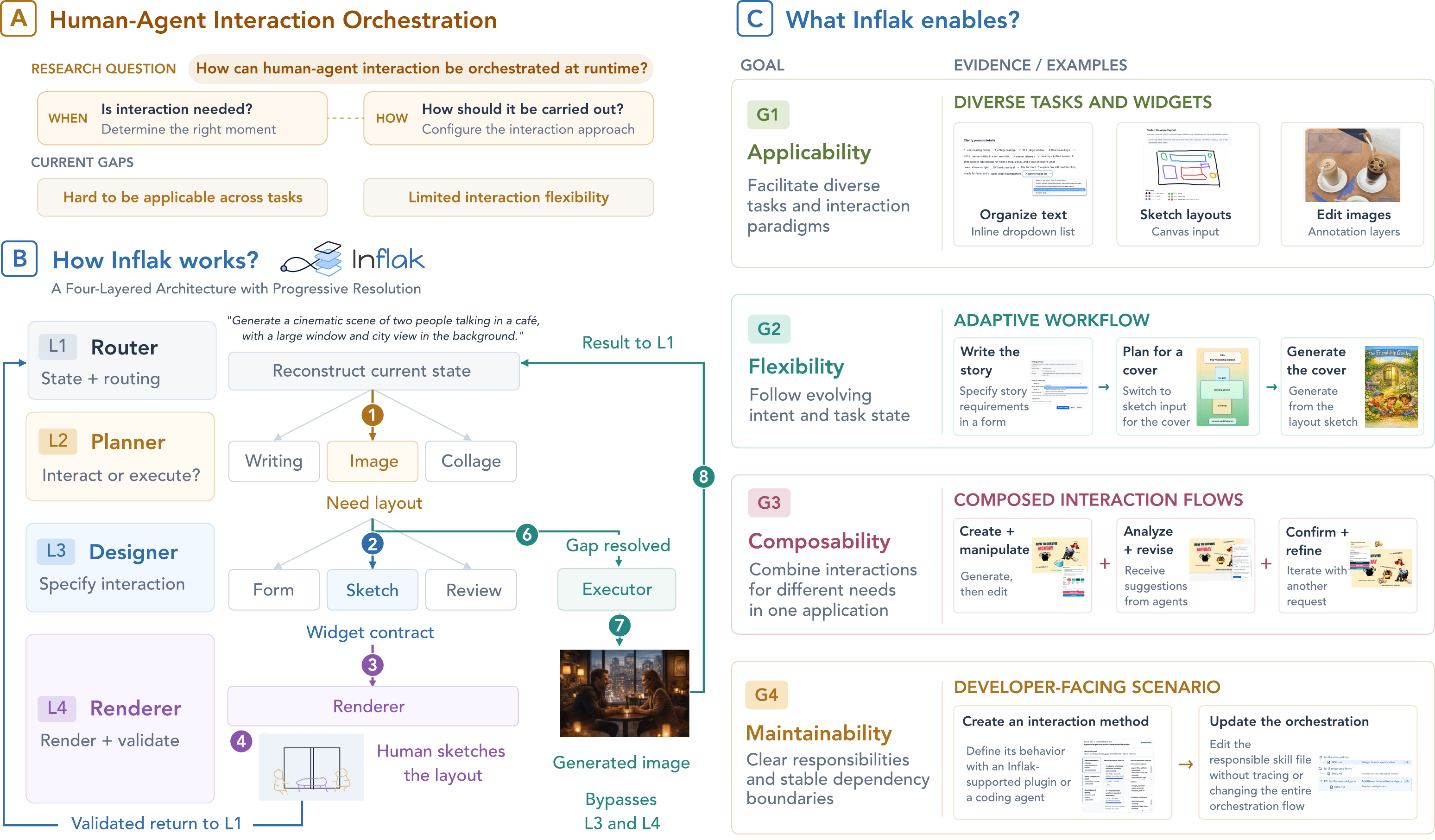}
  \caption{Inflak provides a unified four-layered architecture to orchestrate diverse and dynamic human-agent interaction through progressive resolution, achieving the goals of applicability, flexibility, composability, and maintainability. (A) illustrates the research gap. (B) outlines the architecture of Inflak with an example. (C) demonstrates cases to verify that Inflak can fulfill the four goals. 
  \revised{More information and examples are available on the project website: \href{https://inflak-orchestration.github.io/}{https://inflak-orchestration.github.io/}.}}
  \Description{Three-panel overview of Inflak’s motivation, architecture, and capabilities. (A) Task-specific orchestration is hard to reuse across tasks, while agent-triggered orchestration offers limited interaction methods, leaving rich interaction and cross-task orchestration fragmented. (B) Inflak uses four layers with progressive resolution: L1 Router reconstructs the current state and routes requests; L2 Planner decides whether to interact or execute; L3 Designer specifies an interaction, such as a form, sketch, or review; and L4 Renderer renders and validates it. In the illustrated workflow, a need for layout information prompts a human sketch, which returns to L1 after validation. Once the gap is resolved, the executor generates an image and returns the result to L1, bypassing L3 and L4. (C) Examples illustrate four goals: applicability across organizing text, sketching layouts, annotating images, and exploring data; flexibility to shift from writing a story to generating its cover; composability through creating, analyzing, and refining a poster using different interactions; and maintainability through specifying a widget and adding or editing its L3 skill.}
  \label{fig: teaser}
\end{teaserfigure}


\maketitle

\input{sections/1-introduction}
\input{sections/2-related-work}
\input{sections/4-framework}
\input{sections/5-gallery}
\input{sections/6-tech-evaluation}
\input{sections/7-case-study}

\input{sections/8-discussion}
\input{sections/9-conclusion}
\input{sections/10-acknowledgement}


\bibliographystyle{ACM-Reference-Format}
\bibliography{reference}









\end{document}

%% file: meta/commands.tex
\usepackage{soul}
\newcommand{\revised}[1]{{#1}}

\newcommand{\future}[1]{}

\usepackage{hyperref}
\usepackage[capitalise]{cleveref}
\usepackage{booktabs}
\usepackage{multirow}
\usepackage{booktabs}
\usepackage{tabularx}
\usepackage{multirow}
\usepackage{colortbl}
\usepackage{tikz}
\usetikzlibrary{matrix}
\usepackage{graphicx}
\usepackage{wrapfig}
\usepackage{xcolor}
\usepackage{listings}

\definecolor{stagepurple}{RGB}{204, 204, 255}   
\definecolor{stageorange}{RGB}{255, 204, 153}   
\definecolor{stageblue}{RGB}{173, 216, 230}     
\definecolor{stageteal}{RGB}{139, 204, 139}     
\definecolor{stagegrey}{RGB}{224, 224, 224}     

\definecolor{codeblue}{HTML}{2568EF}
\definecolor{codegreen}{HTML}{057516}
\definecolor{studygreen}{HTML}{61af6c}
\definecolor{studyred}{HTML}{a2362e}

\lstdefinelanguage{Scaffold}{ 
morekeywords=[1]{slide,point,text,create,update,move,remove,feedback,explain,question,hint,instruct,model}, 
morekeywords=[2]{Action,Message,Artifact},
keywordstyle=[1]\color{black}, 
keywordstyle=[2]\color{codegreen},
identifierstyle=\color{black}, 
basicstyle=\ttfamily, 
columns=fullflexible, 
keepspaces=true, 
showstringspaces=false 
}

\lstdefinelanguage{Scaffoldedaction}{
morekeywords=[1]{Action,Scaffolded}, 
keywordstyle=[1]\color{codeblue}, 
identifierstyle=\color{black}, 
basicstyle=\ttfamily, 
columns=fullflexible, 
keepspaces=true, 
showstringspaces=false 
}

\crefname{table}{Tab.}{Tabs.}
\Crefname{table}{Table}{Tables}
\crefname{section}{Sec.}{Secs.}
\Crefname{section}{Section}{Sections}
\Crefname{figure}{Fig.}{Fig.}
\Crefname{figure}{Figure}{Figures}

\usepackage{enumitem}
\usepackage{longtable}

\usepackage{fontawesome}

\usepackage[framemethod=TikZ]{mdframed}
\definecolor{mycolor}{RGB}{139, 143, 190}
\definecolor{bgcolor}{RGB}{242, 243, 249}

\newmdenv[
  backgroundcolor=bgcolor,
  topline=false,
  bottomline=false,
  rightline=false,
  leftline=false,
  skipabove=0.5em,
  skipbelow=0.5em,
  innerleftmargin=0.5em,
  innerrightmargin=0.5em,
  innertopmargin=0.5em,
  innerbottommargin=0.5em,
  singleextra={
    \draw[mycolor, line width=0.5pt] ([yshift=0pt]O) -- ([yshift=0pt]O-|P);
    \draw[mycolor, line width=0.5pt] ([yshift=0pt]P-|O) -- ([yshift=0pt]P);
    \draw[mycolor, line width=0.5pt] ([yshift=1.5pt]O) -- ([yshift=1.5pt]O-|P);
    \draw[mycolor, line width=0.5pt] ([yshift=-1.5pt]P-|O) -- ([yshift=-1.5pt]P);
  },
  secondextra={
    \draw[mycolor, line width=0.5pt] ([yshift=0pt]O) -- ([yshift=0pt]O-|P);
    \draw[mycolor, line width=0.5pt] ([yshift=1.5pt]O) -- ([yshift=1.5pt]O-|P);
  },
  firstextra={
    \draw[mycolor, line width=0.5pt] ([yshift=0pt]P-|O) -- ([yshift=0pt]P);
    \draw[mycolor, line width=0.5pt] ([yshift=-1.5pt]P-|O) -- ([yshift=-1.5pt]P);
  }
]{coloredquotation}

\definecolor{output}{RGB}{222,234,240}
\definecolor{mycoloroutput}{RGB}{177,211,229}
\newmdenv[
  backgroundcolor=output,
  topline=false,
  bottomline=false,
  rightline=false,
  leftline=false,
  skipabove=0.5em,
  skipbelow=0.5em,
  innerleftmargin=0.5em,
  innerrightmargin=0.5em,
  innertopmargin=0.5em,
  innerbottommargin=0.5em,
    singleextra={
    \draw[mycoloroutput, line width=0.5pt] ([yshift=0pt]O) -- ([yshift=0pt]O-|P);
    \draw[mycoloroutput, line width=0.5pt] ([yshift=0pt]P-|O) -- ([yshift=0pt]P);
    \draw[mycoloroutput, line width=0.5pt] ([yshift=1.5pt]O) -- ([yshift=1.5pt]O-|P);
    \draw[mycoloroutput, line width=0.5pt] ([yshift=-1.5pt]P-|O) -- ([yshift=-1.5pt]P);
  },
  secondextra={
    \draw[mycoloroutput, line width=0.5pt] ([yshift=0pt]O) -- ([yshift=0pt]O-|P);
    \draw[mycoloroutput, line width=0.5pt] ([yshift=1.5pt]O) -- ([yshift=1.5pt]O-|P);
  },
  firstextra={
    \draw[mycoloroutput, line width=0.5pt] ([yshift=0pt]P-|O) -- ([yshift=0pt]P);
    \draw[mycoloroutput, line width=0.5pt] ([yshift=-1.5pt]P-|O) -- ([yshift=-1.5pt]P);
  }
]{coloredquotationoutput}

\colorlet{punct}{red!60!black}
\definecolor{delim}{RGB}{20,105,176}
\colorlet{numb}{magenta!60!black}

\lstdefinelanguage{json}{
    basicstyle=\footnotesize\ttfamily,
    numbers=left,
    numberstyle=\color{gray}\footnotesize\ttfamily,
    numbersep=4pt,
    tabsize=2,
    showstringspaces=false,
    breaklines=true,
    xleftmargin=16pt,
    literate=
     *{0}{{{\color{numb}0}}}{1}
      {1}{{{\color{numb}1}}}{1}
      {2}{{{\color{numb}2}}}{1}
      {3}{{{\color{numb}3}}}{1}
      {4}{{{\color{numb}4}}}{1}
      {5}{{{\color{numb}5}}}{1}
      {6}{{{\color{numb}6}}}{1}
      {7}{{{\color{numb}7}}}{1}
      {8}{{{\color{numb}8}}}{1}
      {9}{{{\color{numb}9}}}{1}
      {:}{{{\color{punct}{:}}}}{1}
      {,}{{{\color{punct}{,}}}}{1}
      {\{}{{{\color{delim}{\{}}}}{1}
      {\}}{{{\color{delim}{\}}}}}{1}
      {[}{{{\color{delim}{[}}}}{1}
      {]}{{{\color{delim}{]}}}}{1},
}

\lstdefinelanguage{scaffold in prompt}{
    basicstyle=\footnotesize\ttfamily,
    numbers=left,
    numberstyle=\color{gray}\footnotesize\ttfamily,
    numbersep=4pt,
    tabsize=2,
    showstringspaces=false,
    breaklines=true,
    xleftmargin=16pt,
    morekeywords=[1]{slide,point,text,create,update,move,remove,feedback,explain,question,hint,instruct,model},
    morekeywords=[2]{Scaffolded,Action,Message,Artifact,Event},
    morekeywords=[3]{String,Boolean,Number},
    keywordstyle=[1]\color{black},
    keywordstyle=[2]\color{codegreen},
    keywordstyle=[3]\color{stringcolor},
    identifierstyle=\color{black},
    literate=
     *{0}{{{\color{numb}0}}}{1}
      {1}{{{\color{numb}1}}}{1}
      {2}{{{\color{numb}2}}}{1}
      {3}{{{\color{numb}3}}}{1}
      {4}{{{\color{numb}4}}}{1}
      {5}{{{\color{numb}5}}}{1}
      {6}{{{\color{numb}6}}}{1}
      {7}{{{\color{numb}7}}}{1}
      {8}{{{\color{numb}8}}}{1}
      {9}{{{\color{numb}9}}}{1}
      {:}{{{\color{punct}{:}}}}{1}
      {,}{{{\color{punct}{,}}}}{1}
      {[}{{{\color{delim}{[}}}}{1}
      {]}{{{\color{delim}{]}}}}{1},
}

\definecolor{myblue}{RGB}{47, 85, 151}

\PassOptionsToPackage{svgnames}{xcolor}
\usepackage{xcolor}
\usepackage{xspace, xpunctuate}
\usepackage{listings}  
\usepackage{enumitem}  
\usepackage{multirow}
\definecolor{keywordcolor}{rgb}{0.56, 0.13, 0.00}
\definecolor{ndkeywordcolor}{rgb}{0.05, 0.46, 0.17}
\definecolor{commentcolor}{rgb}{0.41, 0.64, 0.70}
\definecolor{stringcolor}{rgb}{0.25, 0.44, 0.63}
\lstdefinelanguage{TypeScript}{
  keywords={typeof, new, true, false, catch, function, return, null, catch, switch, var, if, in, while, do, else, case, break, boolean},
  morekeywords={[2]{class, export, throw, implements, import, this}},
  identifierstyle=\color{black},
  sensitive=false,
  comment=[l]{//},
  morecomment=[s]{/*}{*/},
  commentstyle=\color{commentcolor}\ttfamily,
  stringstyle=\color{stringcolor}\ttfamily,
  morestring=[b]',
  morestring=[b]"
}
\lstdefinelanguage{mylang}{
}
\lstdefinestyle{mystyle}{
  basicstyle=\footnotesize\ttfamily,
    numbers=left,
    numberstyle=\color{gray}\footnotesize\ttfamily,
    numbersep=4pt,
  frame=none,
  columns=flexible,
  xleftmargin=10pt,
  aboveskip=-1pt,
  belowskip=0pt,
  language=mylang
}

%% file: sections/0-abstract.tex
\begin{abstract}
LLM-powered agents increasingly support open-ended activities, yet orchestrating when and how agents interact with humans remains challenging. Existing approaches often provide interaction mechanisms tied to particular tasks or execution patterns, lacking a unified architecture for diverse and evolving interaction needs. 
We introduce Inflak, a unified four-layer architecture for human–agent interaction orchestration that separates state reconstruction and routing, interaction and execution planning, interaction design, and interface rendering. 
Its layered design with progressive resolution enables general applicability, runtime flexibility, module composability, and orchestration maintainability. 
We implement Inflak as agent-native, skill-based plugins within a general-purpose agent framework and evaluate it through a gallery of various application scenarios and interaction paradigms, technical performance testing, and a comparative implementation case study.
The evaluations show that Inflak can serve as a general and extensible foundation for orchestrating diverse human-agent interaction flows.


\end{abstract}

%% file: sections/1-introduction.tex
\section{Introduction}\label{sec:introduction}

Large language model (LLM)-powered agents are increasingly supporting a wide range of open-ended activities~\cite{pang2025llmification}. 
Along with the increasing capability and applicability, researchers and designers have explored various ways for humans and AI models or agents to interact with each other in efficient and effective ways.
For example, DreamSheets~\cite{Almeda2024dreamsheets} provides a spreadsheet-like interface for users to expand instructions along given dimensions conveniently, such as style and color usage, before task execution.
DynaVis~\cite{baithilingam2024dynavis} applies synthesized widgets to enable users to manipulate generated artifacts in a targeted manner.
A recent work proposed Interaction-Augmented Instruction (IAI)~\cite{shen2026interaction} model to systematically characterize these interaction paradigms.
It presents twelve atomic paradigms and one additional new paradigm, characterizing interactions by their timing, such as before or after invoking AI models, and their targets, including prompts, generated artifacts, and non-verbal information that supplements prompts.
These observations suggest that no single interaction form is suitable for every situation. 
Interaction is often embedded directly in an ongoing activity and grounded in the immediate task state, while the form of interaction can vary substantially and adapt to different task types, user needs, and stages of execution.
Therefore, as the task state evolves, an agent must determine \emph{when} human interaction is needed and \emph{how} it should be carried out, 
while coordinating interactions with autonomous execution. 
We refer to this runtime problem as \emph{human-agent interaction orchestration}. 

Orchestrating human-agent interaction is non-trivial, since the interactions are highly in-situ and dynamic~\cite{li2026situated}.
Also, the agent users and developers may have their special requirements, such as dedicated interaction design to facilitate existing workflow, to be supported by agents at runtime.
These features require designing and implementing orchestration to meet four key pillars: \emph{general applicability} across tasks and runtime capabilities, \emph{flexibility} to initiate and resolve interaction according to evolving task state, \emph{composability} to combine and extend heterogeneous interaction and execution capabilities, and \emph{maintainability} through explicit and separable orchestration responsibilities.
However, these needs are difficult to meet in existing systems, which progressively make interaction decisions from application design toward runtime control. 
A common strategy is to encode workflows in application code and provide predefined interfaces for rich forms of interaction~\cite{chung2022talebrush, wang2024promptcharm}. 
While effective within their intended tasks, both the available interaction forms and when they are used are largely determined at design time, limiting adaptation to changing tasks and states (\textit{insufficient applicability}). 
More recent systems combine code templates with prompt-based generation to synthesize task-specific interaction widgets at runtime~\cite{baithilingam2024dynavis, zhang2026fromwordstowidget}. 
This moves interaction realization into runtime and enables greater in-situ flexibility, but the generation logic and interaction conventions remain tied to individual applications, making them difficult to compose, reuse, or extend across tasks (\textit{insufficient composability and maintainability}). 
More agentic approaches further move interaction control into the agent itself, allowing prompts or agents~\cite{mozannar2025magenticui} to decide when user input should be invoked through mechanisms such as ask-user-question tools~\cite{anthropic2026askuserquestion}. 
Although broadly applicable across tasks, these approaches typically achieve such generality through a small set of generic interaction primitives, such as structured questions or multiple-choice input, limiting the range of interaction semantics and realizations that can be orchestrated (\textit{insufficient flexibility} in interaction methods). 
Taken together, existing approaches separately provide rich predefined interaction, dynamically generated task-specific interaction, or general agent-triggered interaction, but do not jointly support the requirements of increasingly diverse human-agent interaction. 

To fill the gap of existing orchestration methods and meet the four requirements, we introduce Inflak, a unified orchestration architecture for human-agent interaction.
The architecture contains four layers (Router, Planner, Designer, and Renderer), each supporting multiple instances that can be independently defined and integrated into the system.
The previous layer progressively decide which instance in the next layer should be routed to at runtime.
Specifically, based on the comprehensive IAI model, Inflak dynamically constructs the current interaction state as a graph with entities that have been involved in the interaction (e.g., human users, prompt instructions, and artifacts) and links that have been activated (e.g., a prompt was sent to the agent) in the \textbf{Router} layer.
Then, it will pass the state to the corresponding next \textbf{Planner} to decide whether there is a gap between the current state and an optimal state for agent execution (e.g., ambiguous intent) and then what next link and the end entity of the link should be activated (e.g., to ask human to detail image layout with sketching or to generate the image directly).
If a specific interaction is needed to fill the gap, the Planner will pass necessary information, such as the detailed task and gap needs, to the selected \textbf{Designer}, which designs a specific interaction method to address the gap, such as a sketchboard or a composed form with multiple fields.
The Designer will assemble the information from the Planner layer into the interaction specification and pass it to the final corresponding \textbf{Renderer} layer.
The Renderer produces the interaction interface code based on needs, such as HTML, JavaScript, and CSS code for a web-based app, and validates its return to the runtime for Router to plan the following round.
Each instance of one layer is defined with a unified structure with key function introduction, the approach to interpret information from the previous layer, criteria to route to the next layer, and handoff package design to the instances in the next layer.
The layered design enables a dynamic routing across multiple instances of one layer so that different interactions can be initiated for various contexts at runtime (\textit{flexibility}).
It also facilitates adding a new instance, removing one, or modifying one by developers, while keeping the others not affected (\textit{composability} and \textit{maintainability}).
At the same time, rooted in the IAI model, the unified four-layer structure and the consistent structure of information inside each layer instance are disentangled from detailed contexts and specific interaction design. 
It allows that various interactions and tasks can be included and resolved consistently, achieving broader \textit{applicability}.

We instantiate this architecture as plugins to GitHub Copilot\footnote{\url{https://github.com/features/copilot/agents}}, a widely applied agent, in which the instances of four layers are implemented as the agent's skill files.
To demonstrate Inflak can meet the four requirements, we evaluate its applicability through implementing all 13 interaction paradigms reviewed by Shen et al.~\cite{shen2026interaction} as plugins.
Then, we merge ten plugins representing different interaction paradigms into two plugins, \textit{multi-modal co-creation} through a chat-based interface and \textit{visual collage authoring} through a canvas-based interface, demonstrating the composability and the flexibility to apply various in-situ interactions.
We further evaluate orchestration latency and layer compliance using these two composed plugins.
We also showcase the maintainability of Inflak by comparing adding new interaction to the \textit{multi-modal co-creation} plugin with the Inflak architecture or vibe-coded, modularized skill organization. 

To summarize, our main contributions include:
\begin{itemize}[nosep,leftmargin=*]

\item We introduce Inflak, a unified four-layer orchestration architecture for human-agent interaction that separates and progressively resolves control routing, semantic progression, runtime capability binding, and interface implementation over evolving task state.

\item We implement Inflak as plugins to the Copilot Agent, instantiating the architecture through composable skills, explicit handoff contracts, and capability registries.

\item We evaluate Inflak through an application gallery, a technical evaluation, and a comparative implementation case study, examining it against the requirements for applicability, flexibility, composability, and maintainability.

\end{itemize}

%% file: sections/2-related-work.tex
\section{Related Work}
We reviewed existing work to summarize the characteristics of human-agent interaction to understand the necessity of orchestration mechanism (Sec.~\ref{sec:characteristics_hai}).
We also identified how the orchestration is performed in existing work and discussed why they are insufficient to fulfill the requirements led by the characteristics of human-agent interaction (Sec.~\ref{sec:orchestration_hai}).
We note that some interaction design work targets at interactions with general GenAI models instead of strictly defined LLM-powered agents~\cite{mohammadi2025evaluation}, such as work on prompt preparation~\cite{wang2024promptcharm,zhang2026fromwordstowidget}.
However, their interaction and orchestration methods are mostly transferrable to LLM-powered agents.
As a result, we also included them in our related work.
Similarly, we also reviewed a broader scope of human-AI interaction models in Sec.~\ref{sec:rw_interaction_model} to examine how their characterization can facilitate our design of human-agent interaction orchestration architecture.

\subsection{Characteristics of Human-Agent Interaction}\label{sec:characteristics_hai}
The growing capabilities of agents are reshaping how human participation is integrated into interactive tasks~\cite{sapkota2025agents,zou2026survey}. Human–agent interaction is increasingly embedded across different stages of extended activities such as writing~\cite{qin2025toward, zhang2026fromwordstowidget}, data analysis~\cite{wang2024dataformular}, and visual creation~\cite{park2026interaction}. For example, DreamSheets~\cite{Almeda2024dreamsheets} and TaleBrush~\cite{chung2022talebrush} support intent refinement before generation through structured controls and sketch input. WaitGPT~\cite{xie2024waitgpt} exposes an agent’s ongoing data-analysis process so that users can inspect and modify intermediate steps during execution. After generation, systems such as DynaVis~\cite{baithilingam2024dynavis} and AI-Instruments~\cite{riche2025aiinstruments} further support continued refinement of existing artifacts. Other systems involve agents throughout broader task processes, spanning planning, execution, inspection, and revision~\cite{mozannar2025magenticui,feng2026cocoa}. Collectively, these works show that human–agent interaction can occur at multiple points throughout a task rather than at a single, fixed interaction point.

These systems also demonstrate that interaction varies not only in when it occurs, but also in how it is realized to accommodate different task states and artifact types. For example, Intent Tagging~\cite{gmeiner2025intent} introduces interactions tailored to visual and semantic assets. CoLadder~\cite{yen2024coLadder} supports exploratory coding through a node-based interface for code and natural-language prompts. Such systems often design specialized interaction mechanisms around the particular artifact being manipulated. More recent work further explores interfaces that need not be predefined for each artifact or task type, but can instead be adapted at runtime~\cite{lam2026justintime,xu2026duetui}.

Human–agent interaction is therefore increasingly characterized by both interaction diversity and in-situ adaptivity. As runtime state evolves, systems should determine whether interaction is needed, what it should achieve, and how it should be realized.
Inflak addresses this need by progressively resolving these decisions from the evolving runtime state while supporting reusable interaction capabilities across tasks.


\subsection{Orchestration of Human-Agent Interaction}\label{sec:orchestration_hai}

\revised{Human-agent interaction orchestration determines when human participation is needed, what interaction should be initiated, and how it is coordinated with agent execution~\cite{zou2026survey}. Existing systems implement such orchestration through different mechanisms, ranging from application-defined workflows to dynamically generated interfaces and agent-integrated interaction tools~\cite{luera2024survey, mozannar2025magenticui}.}

One common approach is to encode orchestration in application-level workflows and program logic. Such systems organize user interactions, agent operations, and their transitions through predefined application structures. For example, PromptCharm~\cite{wang2024promptcharm} implements multimodal prompt refinement through a task-specific workflow. Cocoa~\cite{feng2026cocoa} supports human-agent co-planning and co-execution within an explicitly designed collaborative process. These systems enable rich task-specific interaction, but when and how interaction occurs is largely predetermined, \revised{limiting the flexibility in deciding interactions at runtime.}

Another approach uses templates and generative mechanisms to realize interactions dynamically. Rather than implementing every interface in advance, these systems leverage runtime context to synthesize or adapt interaction components. For example, DynaVis~\cite{baithilingam2024dynavis} synthesizes widgets for targeted visualization editing. \revised{Malleable Prompting~\cite{zhang2026fromwordstowidget} transforms preference expressions in natural-language prompts into manipulable GUI controls, enabling users to steer LLM generation and inspect the effects of individual preferences on the output.} Related systems similarly use runtime task context~\cite{zhao2026proactiveva,xu2026duetui}, user goals~\cite{lam2026justintime}, or evolving artifact state~\cite{cao2025generative} to generate or adapt task-specific interfaces, providing greater situated flexibility than fully predefined interaction.
\revised{Though more flexible, these methods still depend heavily on the detailed tool-specific implementation, which can hardly be generalized to other applications.}

More recently, agent frameworks and runtimes have incorporated human interaction as part of agent execution. For example, Claude Agent SDK's AskUserQuestion~\cite{anthropic2026askuserquestion} provides a callable mechanism for agents to request structured user input. \revised{AutoGen~\cite{wu2024autogen} integrates humans into multi-agent conversations through configurable user-proxy agents that collect feedback and relay it to other agents.}
By integrating interaction mechanisms into the agent runtime, these approaches allow agents to request human input without prescribing every interaction point in application code. 
However, their built-in interaction mechanisms commonly emphasize generic and rigid interaction methods, such as questions, choices, or approvals, rather than supporting a broad range of interaction forms for flexible and diverse tasks.


\revised{Overall, these methods can hardly achieve both general applicability and flexibility in delivering diverse and in-situ interactions. 
Furthermore, their implementation often distributed the entire orchestration of human-agent interaction across application workflows, interface generation, and agent execution. 
This fragmentation makes it difficult to coordinate and compose diverse interaction mechanisms within a shared runtime structure for easy maintenance and reuse.}
Inflak addresses these issues by resolving interaction needs from the evolving task state at runtime and further separates interaction planning from capability binding and interface realization, allowing different interaction mechanisms to be selected and instantiated within the same orchestration structure.

\subsection{Models for Understanding and Characterizing Human-AI Interaction}\label{sec:rw_interaction_model}

Prior work has modeled human-agent interaction from different perspectives. \revised{Early work examined how control and initiative are distributed between humans and intelligent systems, establishing conceptual foundations for different levels of automation and mixed-initiative collaboration~\cite{parasuraman2000model,horvitz1999principles}. More recent studies extend these perspectives to dynamic human-AI collaboration. Muller and Weisz~\cite{muller2022extending} examine how collaboration changes across different configurations of human and AI initiative, incorporating dynamism and sociality into existing collaboration frameworks. Building on the concept of agency, Zhang et al.~\cite{zhang2025agency} synthesize collaboration patterns and strategies in human-AI co-creation.
Complementing these perspectives, Chappidi et al.~\cite{chappidi2026who} identify recurring archetypes of roles assigned to LLMs in collaborative decision-making, illustrating how AI can contribute to human decisions in different capacities. Together, these studies provide a high-level understanding of how humans and AI share responsibilities, exercise control, and organize collaboration.}

\revised{Other studies examine human-AI interaction more closely through the ways users communicate with and manipulate AI systems. Moving beyond conversational prompting, researchers have investigated how interface affordances, input modalities, and interaction techniques shape users' ability to express and refine their intentions ~\cite{lehmann2024functional,luera2024survey}. These investigations have led to frameworks that organize recurring interaction practices across tasks and contexts. For instance, COFI~\cite{rezwana2023cofi} describes co-creative interaction in terms of how participants contribute, communicate, and manipulate shared products. Other work provides taxonomies of human-LLM and creative GenAI interaction to characterize recurring communication modes and design patterns~\cite{gao2024taxonomy,hu2025designing}. Related work further explores how GUI operations can complement text prompts to express fine-grained intent~\cite{shen2025prompting, riche2025aiinstruments}. 
Among these studies, Shen et al.~\cite{shen2025prompting} characterize who, how, why, when, and what interactions are used to enhance text prompts.
AI-Instruments~\cite{riche2025aiinstruments} demonstrate how prompts can be embodied as reusable, directly manipulable interface instruments. 
These studies offer concrete design knowledge or guidelines for understanding and developing diverse interaction mechanisms, while providing limited structured and operationalizable characterization of human-AI interactions to guide detailed interaction design and implementation.}

\revised{To represent such diverse interactions within a common structure, researchers have further explored how interaction patterns can be constructed from fundamental elements and their relationships. Tsiakas and Murray-Rust ~\cite{tsiakas2024unpacking} develop a semi-formal representation of information exchange between humans and traditional AI systems powered by deep learning techniques, such as recurrent neural networks or interactive reinforcement learning, in which interaction primitives can be composed into recurring patterns.
However, their method provides limited characterization for systems or agents powered by generative AI models that can be instructed with natural language prompts and other GUI interaction techniques, limiting its applicability in our goal to orchestrate human-agent interaction. }
\revised{More recently, Interaction-Augmented Instruction (IAI)~\cite{shen2026interaction} formalizes human-GenAI interaction as an entity-relation graph connecting humans, interactions, prompts, artifacts, and GenAI. By combining these relationships, IAI systematically characterizes diverse interaction paradigms and further supports their composition into more complex forms.}
Unlike broader taxonomies and design spaces, this representation makes the dependencies among interaction components explicit and provides structures that can be instantiated within an execution process, making IAI a suitable basis for studying a unified orchestration mechanism in human-agent interaction. However, IAI itself does not specify how these structures should be operationalized as an evolving agent runtime: the current task state must be interpreted to determine which interaction or execution is needed, abstract relations must be bound to available capabilities, and subsequent execution must be coordinated as state changes. Inflak builds on this representation to provide these runtime orchestration mechanisms.

%% file: sections/4-framework.tex
\section{Inflak: A Four-Layered Architecture to Unify Human-Agent Interaction Orchestration}
As discussed in Sec.~\ref{sec:introduction}, human-agent interactions are dynamic and in-situ, and thus orchestrating these interactions is non-trivial.
This section introduces how we design Inflak to tackle the challenge with a unified four-layer architecture to meet the design goals (Sec.~\ref{sec:goals}).
Then we introduce how we break down the orchestration problem (Sec.~\ref{sec:operationalize_iai}) and tackle it with the four-layer design of Inflak (Sec.~\ref{sec:four_layer_architecture}).

\subsection{Design Goals}\label{sec:goals}
To better accommodate the complexity of human-agent interaction across diverse runtime states, we derive four design goals to guide our orchestration structure design.

\paragraph{G1. Applicability.} A human-agent interaction orchestration architecture should provide a common organization for coordinating heterogeneous runtime participants, including interactions, users, generated artifacts, and agents. This organization should remain applicable across different tasks, interaction paradigms, artifact types, and modalities, rather than requiring a task-specific orchestration structure for each application.

\paragraph{G2. Flexibility.} An orchestration architecture should allow interaction and execution decisions to be resolved in situ from the evolving runtime state. Human responses, artifact changes, tool results, and intermediate execution outcomes should be able to change whether interaction is needed, what interaction should be initiated,  how the interaction results are interpreted, and what the next operation should be.

\paragraph{G3. Composability.} Orchestration capabilities should be independently addable, replaceable, reusable, and combinable. Introducing a new task-specific planner, interaction mechanism, agent, tool, or subworkflow by agent developers should require localized extension rather than reconstruction of an end to end workflow, while existing capabilities should remain reusable across different orchestration contexts.

\paragraph{G4. Maintainability.} The architecture should make different classes of orchestration decisions explicit and identifiable, so that developers can understand how control, interaction, and execution decisions are made. Changes to an orchestration capability should remain localized enough that it can be added, modified, removed, or debugged without tracing or reconstructing the complete workflow.

\input{sections/4-1-design}

\subsection{Defining the Four-Layer Architecture for Progressive Orchestration}\label{sec:four_layer_architecture}
This section introduces the architecture of Inflak with an overview (Sec.~\ref{sec:inflak_overview}) and layer-by-layer explanation (Sec.~\ref{sec:inflak_L1}-Sec.\ref{sec:inflak_L4}).

\subsubsection{Overview}\label{sec:inflak_overview}
We design Inflak as a four-layer architecture to progressively resolve the three problems through L1 \textit{Router}, L2 \textit{Planner}, L3 \textit{Designer}, and L4 \textit{Renderer}.
Besides L1 Router, each layer can have multiple instances to handle various situations.
Each instance will have the identical structure of input, output, and other detailed information to state a modular function.
For example, a L3 Designer instance defines how multi-field forms should be designed, while another instance states what sketchboards will look like.
The previous layer will decide and invoke a specific instance of the next layer with sending the instance an identical handoff package.

Specifically, L1 Router tackles the first sub-problem by collecting context information and pass it to an instance of the Planner layer based on the current task nature.
The Planner layer interprets the contextual information and then determines whether interaction to fill in information gaps between humans and agents or direct task execution is needed, tackling the second sub-problem.
When interaction is needed, an instance of L3 Designer will be invoked to compile the detailed interaction design based on design knowledge and Planner's identified gaps.
This layer addresses the third problem jointly with L4.
L4 Renderer implements the interaction design based on the technical environment where the interaction happens, such as a chat-based web interface.

With the progressive resolution process, Inflak can accommodate various context and interaction needs dynamically (G2).
Also, separating functions across layers and defining multiple instances for one layer ensures that various interaction paradigms can be easily added to Inflak by reusing other layers and instances (G3).
Updating or removing a specific layer instance will only require minimum changes in the previous layer and will not affect other instances in the same or downstream layers (G4).
 
\begin{figure}[htbp]
  \centering 
  \includegraphics[width=\columnwidth]{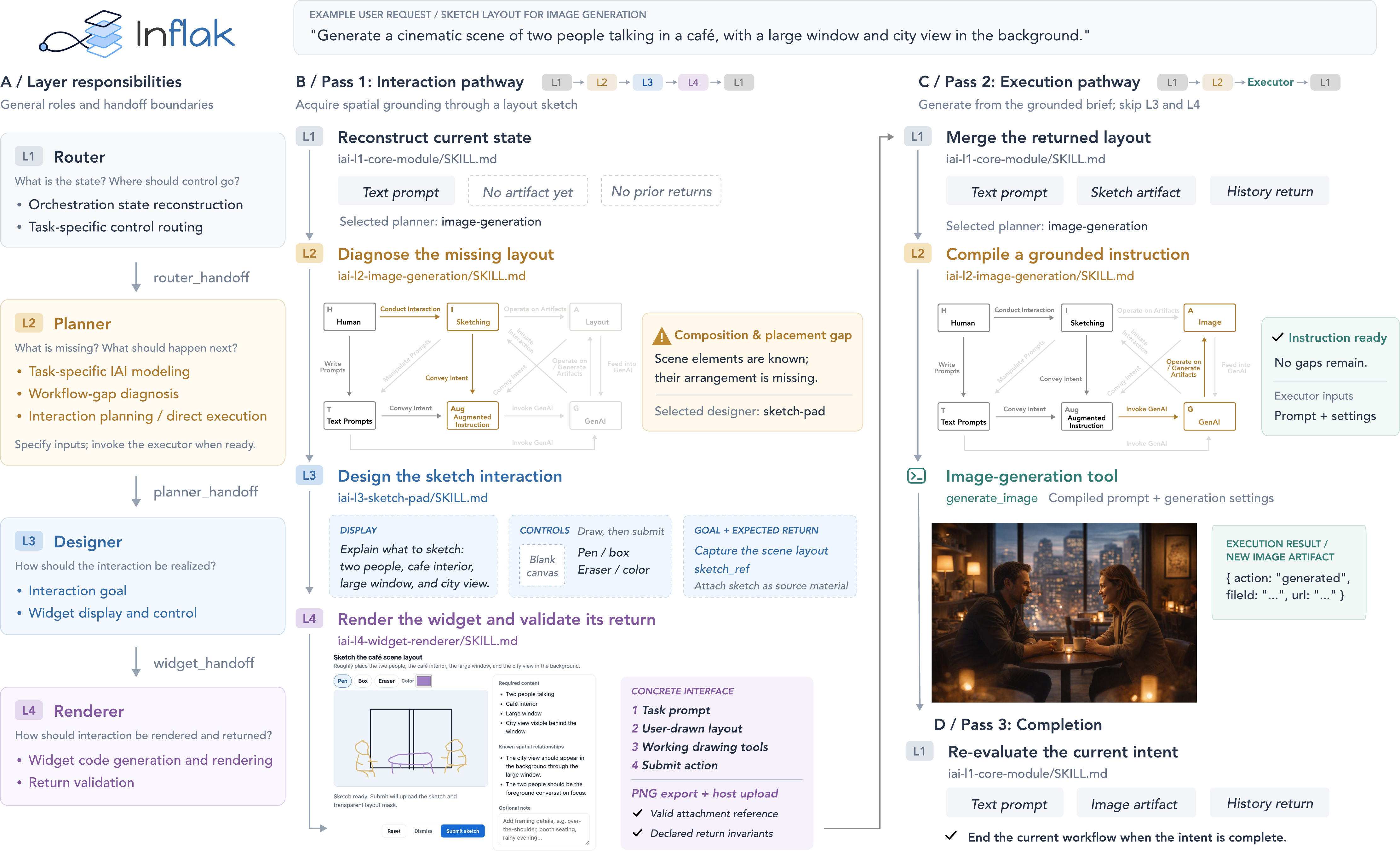}
  \caption{An illustrative sketch-guided image-generation walkthrough in Inflak. (A) General responsibilities of the L1 Router, L2 Planner, L3 Designer, and L4 Renderer. (B) Interaction pathway: L2 identifies a spatial-layout gap, L3 designs a sketch interaction, and L4 renders the widget and validates the user's return. (C) Execution pathway: after L1 incorporates the returned sketch, L2 compiles a sketch-grounded prompt and directly invokes the image-generation tool, bypassing L3 and L4. (D) Completion: L1 incorporates the execution result, re-evaluates the current intent, and ends the active workflow when the intent is satisfied.}
  \label{fig: walkthrough case}
\end{figure}

\subsubsection{L1 Router: Orchestration Context and Control Routing}
\label{sec:inflak_L1}

L1 reconstructs the current orchestration state from the available runtime context, typically including the text prompt, current artifacts and their attributes, and interaction or execution history. It then determines where control should go based on this reconstructed state.
For task-specific routing, L1 consults a routing table formed from the registered L2 planners and their applicability conditions. These conditions may consider the current task intent, artifact type, artifact availability, and other state attributes to determine which planner can meaningfully advance the task. In Fig.~\ref{fig: walkthrough case}, for example, the initial request expresses an image-generation intent while no artifact is yet available, so L1 routes control to the L2 image-generation planner. If no registered planner applies, control is returned to the host agent. If the active task has already been completed, as after the generated image is returned in Fig.~\ref{fig: walkthrough case}, L1 instead terminates the current session.
Every widget return and execution result re-enters L1 and updates the orchestration state before another routing decision is made. This re-entry allows L1 to determine, from the latest state, whether the task should proceed to another planning step, return to the host agent, or end. L1 therefore serves as the runtime boundary for state reconstruction and control routing.

\subsubsection{L2 Planner: Interaction Requirement Planning}
\label{sec:inflak_L2}

L2 interprets the orchestration state through a task-specific IAI model, diagnoses the material gap that currently blocks progress, and plans the next interaction or execution step. Whereas L1 determines which planner should handle the task, L2 determines what that task requires next and why.

The planner binds the core IAI nodes and their relationships to the task domain. In the cafe example (Fig.~\ref{fig: walkthrough case}), the human user (\texttt{H}) provides a scene request (\texttt{T}); sketching constitutes the interaction (\texttt{I}); the resulting layout sketch becomes a grounding artifact (\texttt{A}); and the augmented instruction (\texttt{Aug}) combines the request with the spatial information needed by the image-generation model (\texttt{G}). These bindings allow the planner to reason about both the available information and the dependencies that remain unresolved.

L2 applies a task-specific gap policy to determine whether a gap remains that warrants further human interaction. If interaction is needed, the planner produces an interaction requirement describing the gap to resolve, the interaction goal, the relevant IAI relation and context, and the expected return. If the state is sufficiently specified for autonomous execution, it instead produces an execution requirement describing the operation that should be performed and the task information that should be carried forward.
In the first pass of Fig.~\ref{fig: walkthrough case}, L2 identifies a composition and placement gap and emits an interaction requirement whose goal is to obtain a rough spatial layout from the user, with a layout artifact as the expected return.
When the submitted sketch later re-enters the orchestration loop, L2 reassesses the updated state and emits an execution requirement to generate the image using the original scene request together with the returned layout.

\subsubsection{L3 Designer: Interaction Specification}
\label{sec:inflak_L3}

L3 designs the concrete interaction required by the plan from L2. Given an interaction requirement and its task context, it specializes the intended interaction into a semantic widget contract. The contract refines the interaction goal, determines what information should be presented to the user, specifies the supported interactive controls, and defines the expected return. L3 determines the interaction design without implementing the concrete interface.

In Fig.~\ref{fig: walkthrough case}, L2 identifies the need to obtain the user's intended spatial layout. L3 turns this requirement into a sketching widget specification: the scene request and relevant entities are presented as context, a drawing canvas and appropriate sketching controls support spatial input, and the completed sketch is defined as the expected return. The resulting widget contract is then passed to L4 for concrete interface implementation.

\subsubsection{L4 Renderer: Interface Realization and Return Validation}
\label{sec:inflak_L4}

L4 compiles the renderer-independent widget contract into an executable and sandboxed interface instance, and binds submission behavior to return collection. When an interaction requires a particular presentation style or complex interface behavior, L4 may incorporate reference HTML, CSS, or JavaScript assets while preserving the semantic contract defined at L3.
L4 also validates returns against the expected-return schema and routed back to L1.
In Fig.~\ref{fig: walkthrough case}, L4 realizes the sketch contract as a sandboxed canvas with the required task context, drawing controls, and submission actions. Once submitted, the sketch is exported as a layout artifact and returned in the contract-defined structure after validation. The validated return then re-enters L1 for state reconstruction and further routing.

\subsubsection{Runtime Resolution Pathways}\label{sec:runtime pathways}

Inflak does not require every runtime event to traverse all four layers. 
Instead, the current orchestration state is resolved through different runtime pathways depending on whether further human interaction is required, autonomous execution is ready, or no further Inflak processing is needed. 
Each interaction or execution result subsequently re-enters L1, where the updated state is reconstructed and a new pathway is determined.

\paragraph{Interaction Pathway.}
When L2 identifies a gap that requires human input, the orchestration proceeds through the interaction-design and rendering layers:

\[
L1 \rightarrow L2_{\mathrm{interaction}}
\rightarrow L3
\rightarrow L4
\rightarrow \mathrm{Human\ Return}
\rightarrow L1.
\]

L1 first routes the current state to the applicable task planner. 
L2 identifies the gap to resolve and specifies the corresponding interaction requirement. 
L3 refines this requirement into a task-specific widget contract, and L4 implements the contract as a concrete interface and collects the validated return. 
The returned interaction result then re-enters L1 as part of the updated orchestration state. 
In Fig.~\ref{fig: walkthrough case}B, this pathway is used to obtain the user's intended spatial composition through a sketch.

\paragraph{Execution Pathway.}
When L2 determines that the current state is sufficiently specified for autonomous execution, the interaction-design layers are bypassed:

\[
L1 \rightarrow L2_{\mathrm{execution}}
\rightarrow \mathrm{Executor}
\rightarrow L1.
\]

L2 produces the execution-ready instruction and invokes the corresponding task executor with the required inputs. 
The resulting artifact or execution result returns to L1, where it becomes part of the reconstructed state. 
In the second traversal of Fig.~\ref{fig: walkthrough case}C, the returned layout sketch resolves the spatial gap, allowing L2 to invoke image generation using the request and layout artifact.

\paragraph{Host-Agent Return and Completion.}
L1 may also determine that no further Inflak processing is required. 
If the current event falls outside the responsibility of the registered task planners, control is returned to the host agent. 
If the active task has completed and no further interaction or execution is required, the current orchestration workflow ends:

\[
L1 \rightarrow
\begin{cases}
\mathrm{Host\ Agent},\\
\mathrm{End}.
\end{cases}
\]

These pathways allow the runtime to alternate between human interaction and autonomous execution according to the evolving task state, rather than committing to a fixed interaction sequence in advance.

%% file: sections/4-1-design.tex
\subsection{Formulating the Orchestration Workflow}\label{sec:operationalize_iai}
To achieve the four goals with Inflak, the foundation is a model of human-agent interaction with descriptive power to characterize various interaction paradigms, discriminative power to differentiate those paradigms' usage, and generative power to accommodate potential new paradigms~\cite{Liao2023, Beaudouin-Lafon2004}.
Only after defining interactions between humans and agents in a universal way, it is possible to design a unified orchestration mechanism to scaffold agents in deciding when they should interact with humans and how the interaction would be conducted.
A model with the three powers ensures that Inflak can be generally applicable to existing and also future human-agent interaction (G1).

\begin{figure}[]
  \centering 
  \includegraphics[width=\columnwidth]{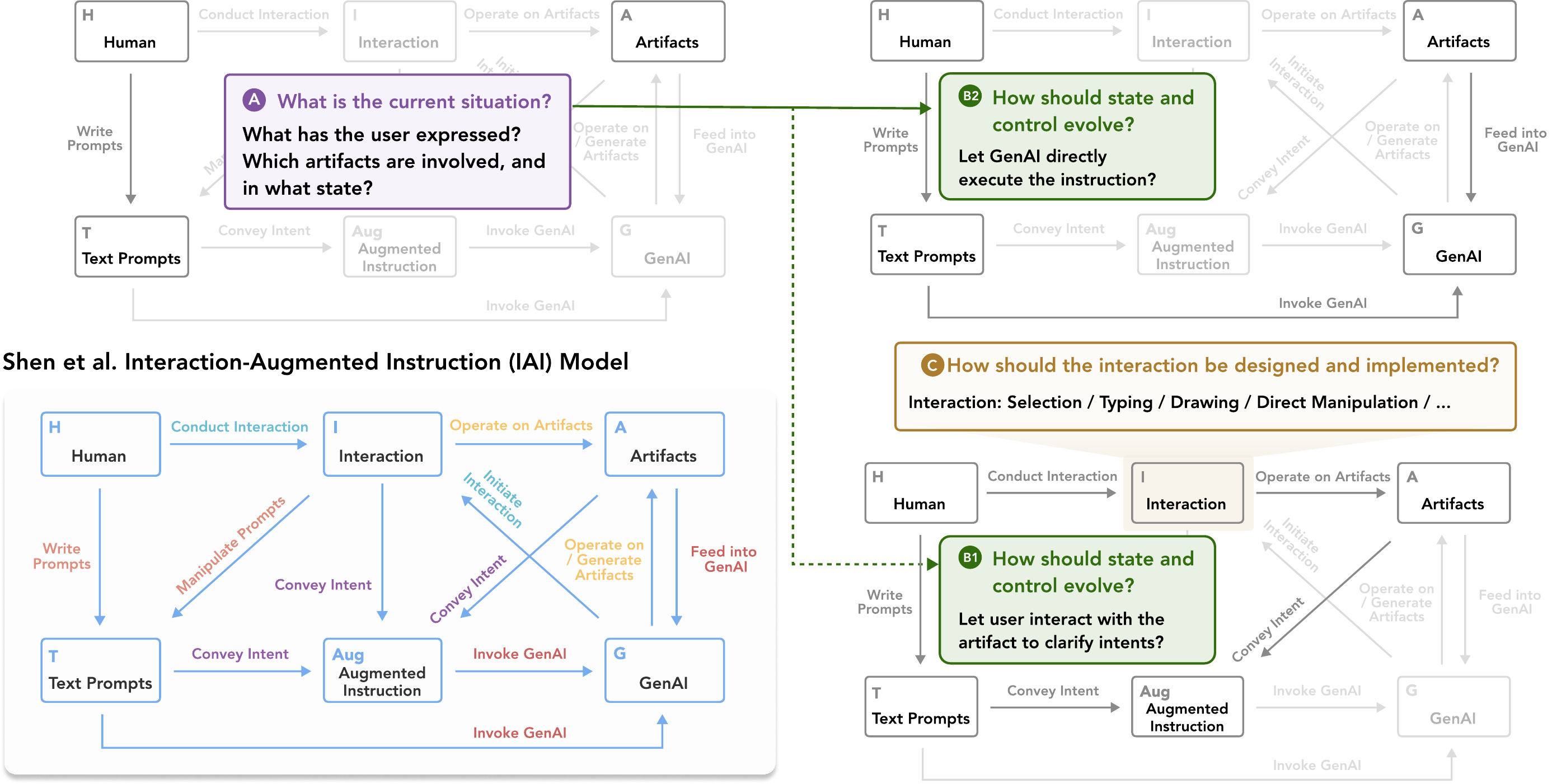}
  \caption{Formulating the orchestration workflow of Inflak based on the IAI model. We derive three concrete sub-problems for orchestration: (A) contextual state interpretation, (B) stateful runtime coordination, and (C) detailed next-step interactions. }
  \label{fig: comparison with IAI}
  \vspace{-0.45cm}
\end{figure}

Based on our survey of human-agent interaction or broadly human-GenAI interaction models (Sec.~\ref{sec:rw_interaction_model}), we identified Interaction-Augmented Instruction (IAI) model as an appropriate starting point, as it captures a broad range of recurring and composable interaction patterns and provides a structured representation amenable to runtime execution. 
IAI model defines a graph consisting of entities and relations to show how information flows among them in human-GenAI interactions (Figure~\ref{fig: comparison with IAI}).
The entities define what participants are in the current interaction, including human, interaction, prompt, interaction-augmented instructions (i.e., a combination of both verbal and non-verbal information produced by interactions as input to GenAI), artifacts, and GenAI.
The relations specify who conducts the interaction and what the interaction targets. 
For example, linking a human user node, an interaction node, and an artifact node means that the user conducts interactions on the artifact.
With the design, interaction paradigms for various scenarios and goals can be derived by combining multiple relations, such as one paradigm to show how interactions on artifacts can be used to express multi-modal intent (an example tool using the paradigm is DirectGPT~\cite{masson2024directgpt}) and another to delineate how generative widgets allow users to manipulate artifacts (an example tool using the paradigm is DynaVis~\cite{baithilingam2024dynavis}).
With the IAI model and its derived interaction paradigms, we abstract and divide our orchestration problem into three sub-problems which need to be solved sequentially:


\paragraph{Build contextual interpretation of IAI entities and relations for decisions.}
IAI paradigms describe the entities involved in an interaction, how these entities are connected, and how interaction is organized in relation to other elements of the paradigm. 
In a runtime setting, however, these descriptive structures must be interpreted against a concrete and continuously evolving task state. 
The system first needs to collect interaction relevant information from sources such as user prompt, artifacts, prior interaction results, and execution outcomes, and use this information to determine how the current runtime state instantiates an abstract interaction structure in the next step. 
This motivates a runtime orchestration flow that first establishes the current context and routes control accordingly (Fig.~\ref{fig: comparison with IAI}(A)), before identifying the next operation needed to advance the task.

\paragraph{Decide IAI paradigm application and task execution through stateful runtime coordination.}
IAI identifies twelve atomic interaction paradigms and demonstrates how the conceptual model can support the derivation of additional paradigms and the composition of multiple paradigms into richer interaction workflows. 
In an agent runtime, however, composition alone does not determine how such a workflow should unfold over time. 
Runtime execution introduces stateful dynamics and lifecycle management concerns. 
The system must determine which interaction paradigm or execution step is currently applicable and then reason over the corresponding entities and relations to determine what information or state change is expected before the task can progress. 
It includes deciding whether human interaction is currently needed to address issues such as underspecified intent or unresolved artifact references (Fig.~\ref{fig: comparison with IAI}(B1)).
If needed, it should also select the appropriate interaction paradigm to proceed with, such as asking the user to mark their intent on artifacts.
If no interaction is needed, it should hand over the task to executors, such as a GenAI model (Fig.~\ref{fig: comparison with IAI}(B2)).
Therefore, composing interaction paradigms at the conceptual level must be extended into a stateful coordination process that continuously and progressively maintains and updates the task context throughout runtime execution (G2 and G3).

\paragraph{Detail next-step interaction represented with abstract IAI entities and relations to concrete interaction designs.}
IAI intentionally represents entities and relations at an abstract level, which enables recurring interaction paradigms to be identified and compared independently of particular implementations. 
Runtime orchestration, however, must translate these abstract roles into concrete capabilities that can actually be invoked in the current environment. 
Specifically, the interaction entity $I$ captures the functional role of human interaction without prescribing how that interaction should be performed. Its concrete realization may vary according to the task, modality, artifact, and intended information return, such as selecting among recommended options, providing textual input, sketching over a visual artifact, or directly manipulating an object. 
Runtime orchestration therefore requires a further resolution step that binds abstract interaction roles to appropriate interaction designs and also implement designs as executable interfaces (Fig~\ref{fig: comparison with IAI}(C)).

%% file: sections/5-gallery.tex
\input{sections/5-0-implementation}

\section{Application Gallery}\label{sec: gallery}
In this section, we use an application gallery to examine the applicability, flexibility, and composability through the implementation method introduced in Sec.~\ref{sec:instantiation}. 
Sec.~\ref{sec:gallery_paradigms} demonstrates the general applicability of Inflak through implementing all 13 interaction paradigms summarized by Shen et al.~\cite{shen2026interaction} as plugins (P1-P13).
Then, we showcase how Inflak fulfills the flexibility and composability goal through merging six plugins (P1-6) into one \textit{multi-modal co-creation} plugin (Sec.~\ref{sec: gallery case 1}) and another four plugins (P8, P10-12) into a \textit{visual collage authoring} plugin (Sec.~\ref{sec: gallery case 2}).
Both of them work well to guide agent to orchestrate different interactions dynamically.

\subsection{Main Gallery}\label{sec:gallery_paradigms}
\begin{figure}[htbp]
  \centering 
  \includegraphics[width=\columnwidth]{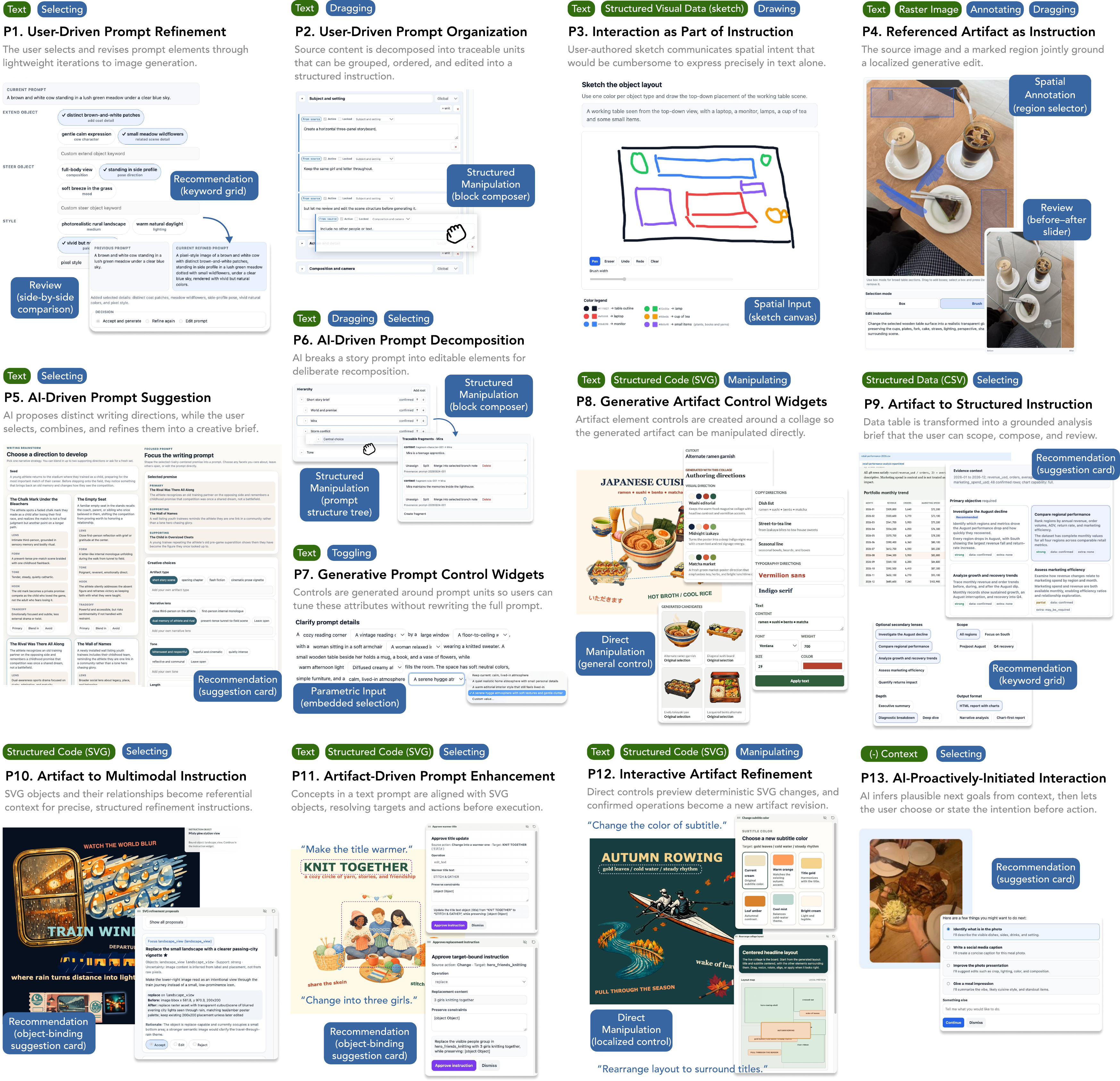}
  \caption{Overview of the Inflak application gallery. The gallery instantiates the thirteen atomic interaction paradigms (P1-P13) across diverse artifact substrates, including text, sketches, raster images, structured data, and vector graphics, and interaction realizations, including selection, dragging, drawing, annotation, toggling, and direct manipulation.}
  \label{fig: gallery overview}
\end{figure}

These application cases capture the functional role of human participation in the task flow and are grounded in all twelve atomic IAI interaction paradigms (P1-P12) extracted from more than 60 GenAI tools and a newly identified paradigm (P13) by the authors of IAI~\cite{shen2026interaction}.
Fig.~\ref{fig: gallery overview} provides an overview of these cases and their coverage. Demonstration videos and additional information for all gallery cases, including their interaction realizations and artifact substrates, are available on the companion website\footnote{\url{https://inflak-orchestration.github.io/Inflak-gallery/}}.

These implemented applications capture different interaction strategies for various tasks.
Basic selection-based widgets enable agents to present necessary information for users' easy selection in an in-situ manner based on the current task, artifact, and timing (P1, P5, P9, P13).
More diverse interaction techniques are also supported by Inflak, allowing humans and agents to communicate essential information conveniently.
Draggable blocks let users organize and recompose individual units (P2, P6).
Embedded toggle widgets allow users to adjust prompt attributes in place without rewriting the full prompt (P7). 
Sketch-based interaction (P3, P4) lets users communicate spatial intent through the placement and relative arrangement of drawn elements.
The sketch itself becomes part of the instruction, without requiring users to translate these spatial relationships into verbal descriptions. 
Region annotations (P4) ground instructions in selected parts of an image.
Object-bound suggestion widgets (P10, P11) and localized controls (P12) enable targeted refinement of individual elements within multi-object artifacts.
Beyond widget design, these applications also cover different artifact types, including text (e.g., P1-P4), visual (e.g., P8, P11), and data (e.g., P9), and interaction environment, including chat-based interface (e.g., P2, P13) and canvas-based interface (e.g., P11, P12).
The scope of interaction design that can be supported by Inflak justifies that it is a generally applicable orchestration architecture.


\newpage

\subsection{Case I: Multimodal Co-Creation}\label{sec: gallery case 1}

\begin{figure}[htbp]
  \centering 
  \includegraphics[width=\columnwidth]{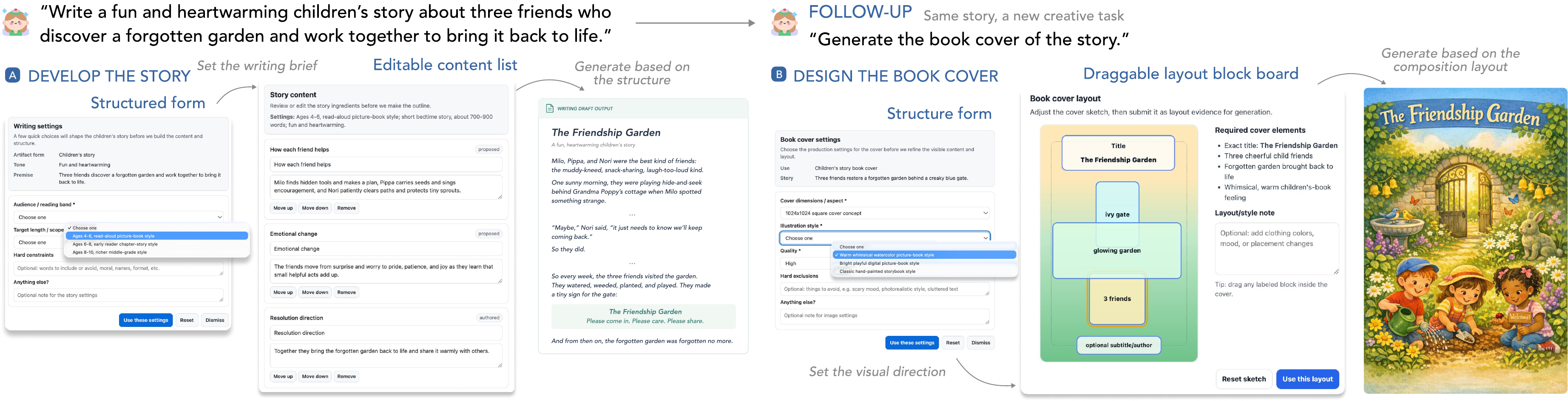}
  \caption{Cross-task multimodal co-creation with Inflak. (A) A writing task transitions into (B) a related book-cover design task, involving different artifact types and task-specific interaction mechanisms while reusing common widgets, such as structured forms, across task boundaries.}
  \label{fig: gallery main}
\end{figure}

General-purpose agents often support sequences of related but heterogeneous tasks rather than a single predefined workflow. We instantiate this scenario with the multimodal co-creation task (Fig.~\ref{fig: gallery main}), where the user first develops a children’s story and then follows up with a new creative task for a book cover. In the writing task, the user specifies the writing brief through a structured form and refines the planned story content through an editable content list before generation. The follow-up task shifts from text authoring to visual creation, where a structured form specifies the visual direction and a draggable layout board allows the user to organize the cover composition before image generation.

The two tasks therefore involve different artifact types, task-specific reasoning, and interaction realizations, while remaining connected through the same evolving activity. Inflak organizes these heterogeneous capabilities as separate task-specific skills under a shared orchestration structure, while allowing interaction widgets to be reused and recombined across task boundaries. For example, the same structured-form widget is reused to capture task-specific settings in both writing and visual design, while each task composes it with specialized interaction mechanisms suited to its artifact and workflow. The runtime can therefore route changing user intent and artifact state to the appropriate task planner and interaction mechanism without requiring an entirely separate interaction stack for each task. 

\subsection{Case II: Collage Authoring}\label{sec: gallery case 2}

Many authoring tasks do not follow a single interaction pattern, but require different forms of interaction as the artifact evolves. 
We examine this setting through an SVG collage-authoring task that combines the P8 and P10–P12 interaction paradigms in Fig.~\ref{fig: gallery overview} into an iterative authoring process around a persistent artifact.

\begin{figure}[htbp]
  \centering 
  \includegraphics[width=\columnwidth]{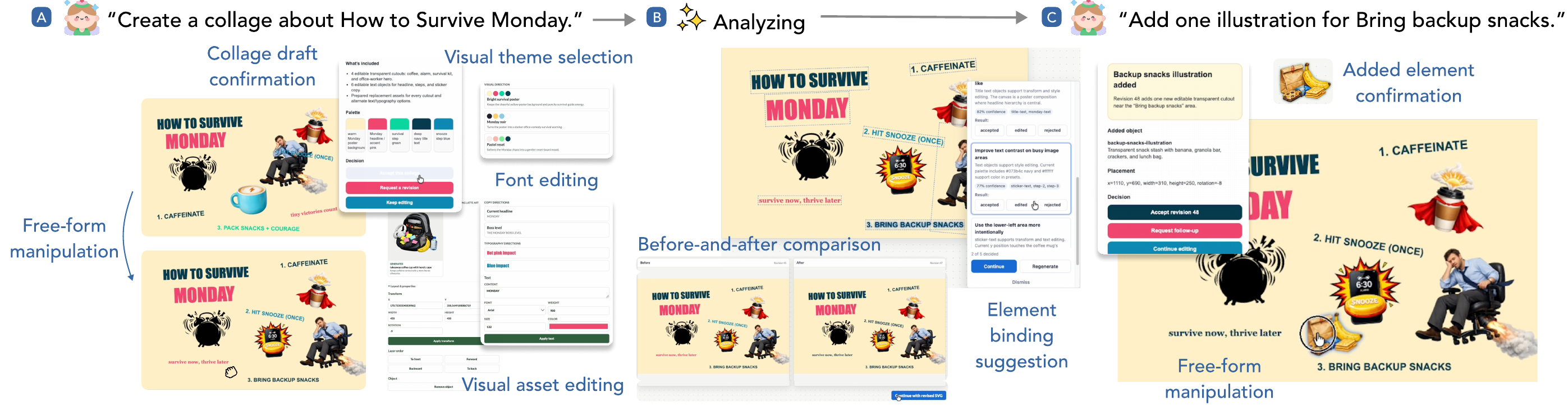}
  \caption{Iterative collage authoring with Inflak. Different interaction flows are composed around the same persistent SVG artifact as the task progresses from (A) initial creation, to (B) analysis and revision, and to (C) incremental refinement request.}
  \label{fig: gallery collage}
\end{figure}

As shown in Fig.~\ref{fig: gallery collage}, the workflow begins with an open-ended collage request and progressively introduces different interaction mechanisms according to the current authoring stage. During initial creation, the user can confirm a generated collage draft, select a visual theme, edit fonts and visual assets, or directly manipulate the collage. Once an artifact has been established, subsequent interactions operate on its current state: the system can analyze the composition, suggest bindings between elements, and provide before-and-after comparison for revision. The workflow can then continue with incremental requests, such as adding a new illustration, where the newly generated element is confirmed and incorporated into the existing collage before further manipulation.

These interaction flows are organized through composable authoring skills rather than encoded as a single fixed workflow. Artifact management maintains the evolving collage and its associated visual assets across interaction rounds, allowing different skills and interaction paradigms to operate on the same persistent artifact. This case therefore illustrates how Inflak composes heterogeneous interaction flows across successive stages of a stateful task, enabling a more complete iterative workflow while preserving continuity of the underlying artifact.

%% file: sections/5-0-implementation.tex
\section{Implementing Inflak as Agent Plugins} \label{sec:instantiation}

We implement Inflak as skill-based agent plugins to GitHub Copilot, a widely applied agent.
In a plugin, each layer defines a fixed runtime responsibility and layer instances define concrete capabilities through skills and shared handoff contracts.
Through integrating the plugin with Copilot SDK\footnote{\url{https://github.com/github/copilot-sdk}} as backend, the agent can leverage Inflak to dynamically interact with human users through a web-based interface.
We choose skills as the implementation unit because they provide a lightweight and agent-native mechanism.
It allows modular behavior without modifying the host agent's core execution logic. 
Each skill can independently encode a layer-specific capability, its applicability, and its expected inputs and outputs in a readable specification, while remaining dynamically invocable by the host agent at runtime. This modular structure maps naturally to Inflak's layered design, allowing individual layer capabilities to be implemented and extended independently while coordinating through shared handoff contracts.
In the following, we describe how the four-layer architecture is implemented and present the resulting plugins in Sec.~\ref{sec: gallery}.
\revised{Our implementation is available at \url{https://github.com/Inflak-orchestration/Inflak-demo}.}


\begin{figure}[htbp]
  \centering 
  \includegraphics[width=0.9\columnwidth]{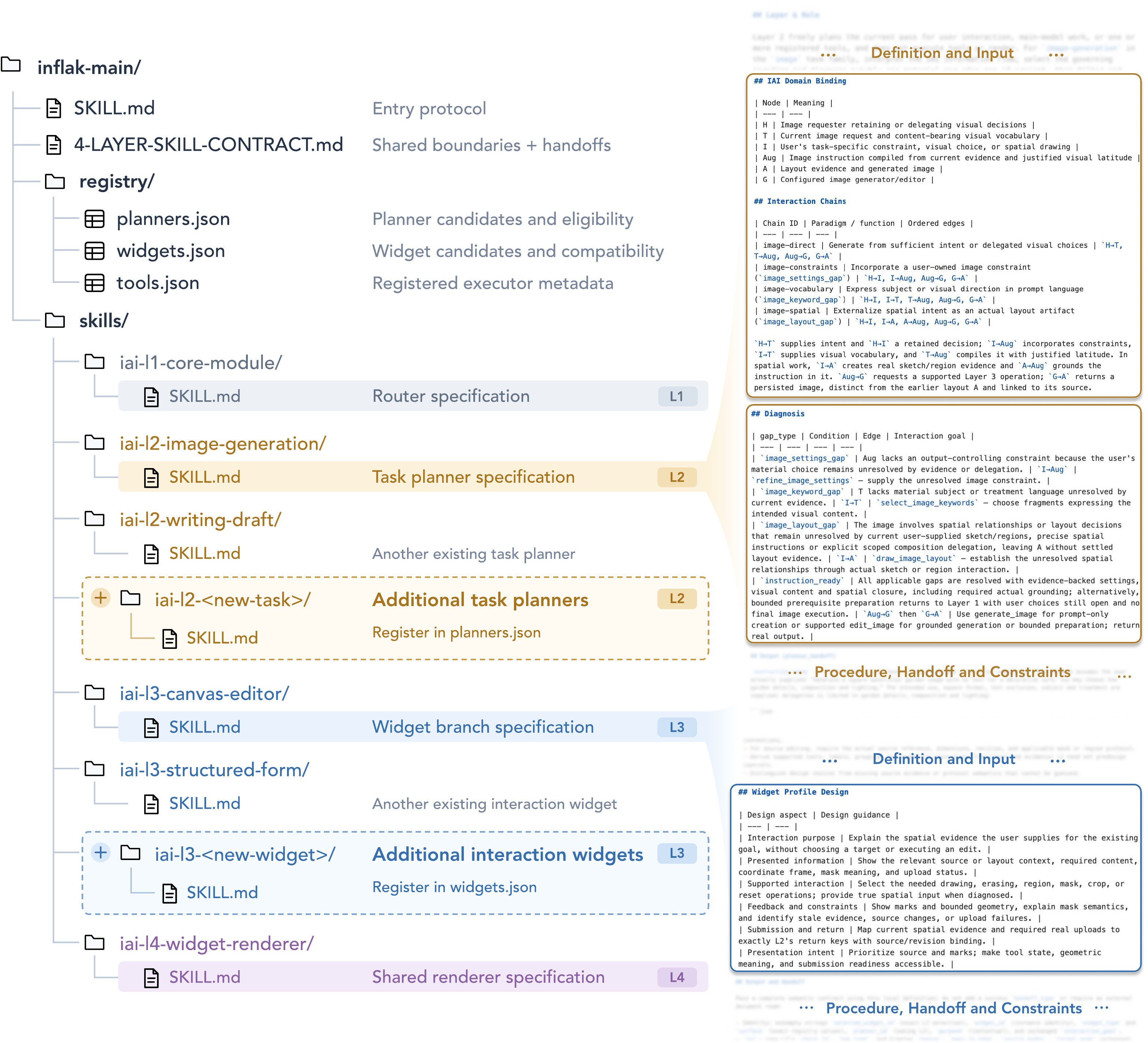}
  \caption{File organization of the Inflak agent plugin example. Shared contracts specify layer boundaries and handoffs, while registries describe available planners, interaction widgets, and executors. Layer-specific skills provide behavioral instances for orchestration and interaction realization. Dashed nodes indicate extension points for additional task planners (L2 Planner) and interaction widgets (L3 Designer). The right-side snapshots show representative specifications of an L2 Planner of image generation task and an L3 Designer of canvas editor.}
  \label{fig: plugin file structure}
  \vspace{-0.45cm}
\end{figure}

\subsection{Skill-Based Plugin Organization} 

The four layers are instantiated through different classes of skills. L1 is implemented as a shared router skill that reconstructs the current orchestration state and selects an eligible L2 planner. L2 consists of task-specific planner skills that determine what needs to happen next. L3 consists of reusable interaction skills that realize an interaction requirement as a concrete widget specification, while L4 is implemented as a shared renderer skill responsible for interface realization and structured return collection. The root \texttt{SKILL.md} defines the entry protocol, while \texttt{4-LAYER-SKILL-CONTRACT.md} specifies the shared boundaries and handoff structure across these skills.
Adjacent layers communicate through explicit handoff objects rather than unconstrained agent traces. Each handoff carries the information required by the next layer while preserving decisions already made upstream. 
Figure~\ref{fig: plugin file structure} presents an overview of the skill organization and key parts of L2 and L3 skills.
A full plugin example is available in the supplemental material.

A plugin also includes three readable registries. As shown in Figure~\ref{fig: plugin file structure}, 
\texttt{planners.json} describes candidate L2 planners and their eligibility conditions, 
\texttt{widgets.json} describes reusable L3 interaction capabilities and their compatibility constraints, and \texttt{tools.json} records metadata for task executors. 
These files allow agents and human developers to discover and understand the capability of one Inflak plugin.

\subsection{Developing a Plugin}\label{sec:authoring_plugin}

Inflak provides shared contracts that specify the expected responsibility, required content, and handoff structure of each layer instance. 
These contracts serve as authoring guidance for translating task and interaction descriptions into framework-compatible skill files. 
Besides manually authoring them by following guidance, it is possible to define a new L2 Planner skill in natural language and ask coding agents, such as GitHub Copilot, to implement it. 
For example, a task such as ``select relevant HCI papers every day and produce a daily paper report'' can be translated into an L2 Planner that applies the IAI model to identify task states in which missing information or unresolved decisions constitute an interaction gap. 
L3 Designer instances allow similar flexibility that developers can describe an intended interaction profile in natural language or provide reference implementation code as a template. 
Therefore, a basic Inflak plugin can be authored largely through natural-language interaction with coding agents, based on the contracts constraining the resulting skills into the expected four-layer structure.

\begin{figure}[htbp]
  \centering 
  \includegraphics[width=\columnwidth]{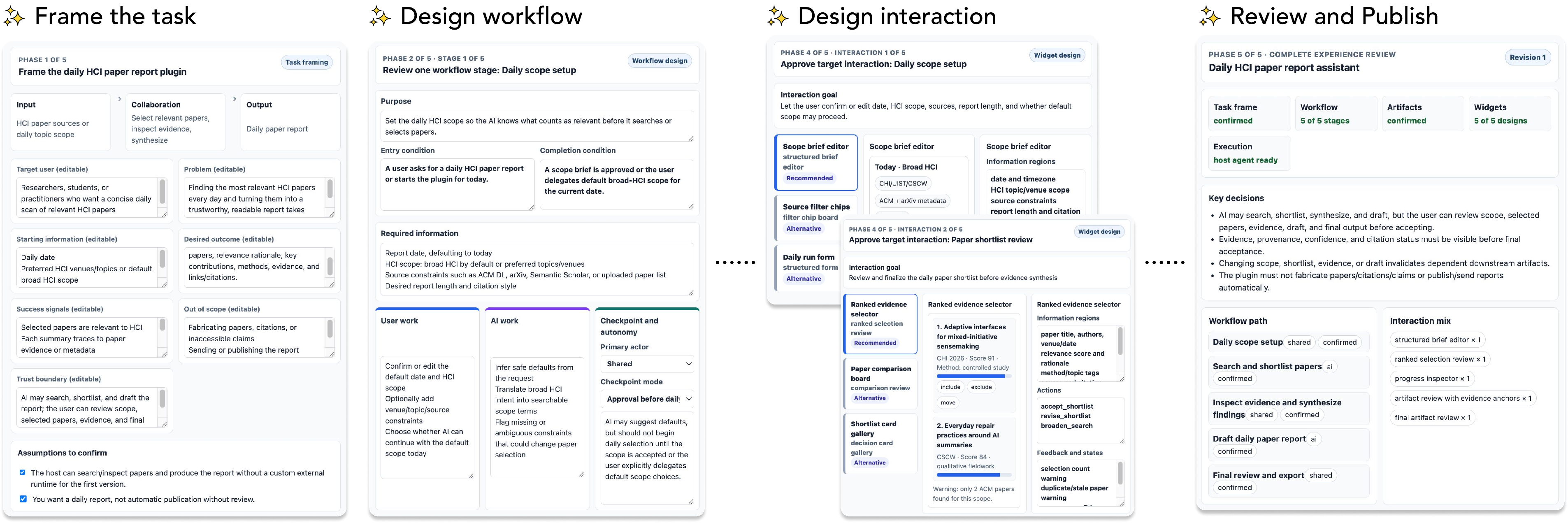}
  \caption{Using \textit{Inflak authoring} plugin to create interactive widgets for daily paper report generation. Based on Inflak's architecture, this plugin supports the authoring process from framing the task, decomposing and designing the workflow, and specifying widget-based interactions, to layer-specific skill files and corresponding registry files.}
  \label{fig: inflak authoring}
  \vspace{-0.45cm}
\end{figure}

We also developed an \textit{Inflak authoring} plugin based on Inflak's four-layer structure.
It can help new developers to get familiar with what information should be provided for a basic Inflak-based plugin by dynamically scaffolding developers with Inflak-orchestrated interactions.
During authoring, the plugin continuously reasons about what information is still missing, when user input is needed, and which interaction mechanism is appropriate for eliciting or aligning that information. 
It can therefore decide, for example, when to ask the user to clarify a task boundary, select among candidate workflow stages, or compare alternative widget designs, rather than following a fixed authoring wizard.
For the daily-paper example (Fig.~\ref{fig: inflak authoring}), the authoring plugin may progressively decompose the task into stages such as confirming the date range, specifying research topics, and filtering candidate papers. 
At each stage, Inflak determines whether an interaction gap exists and, if so, selects an appropriate interaction form, such as a date selector, keyword selection, or paper-filtering cards, to align the emerging plugin design with the user's intent. 
Once the task logic and interaction designs are sufficiently specified and confirmed, the accumulated intent is represented as a plugin blueprint. This blueprint is then translated, according to the shared contracts, into concrete L1-L4 instances and corresponding registry files. 
In this sense, the authoring plugin makes Inflak self-applicable: the same orchestration structure used by downstream plugins is also used to coordinate how those plugins are authored.


\subsection{Extending a Plugin}

The plugin organization provides explicit extension points for adding new task and interaction capabilities. As illustrated by the dashed nodes in Fig.~\ref{fig: plugin file structure}, adding a new task-specific planning capability primarily requires introducing a new L2 Planner skill and registering it in \texttt{planners.json}. Its task scope and eligibility conditions make the new capability visible to the shared L1 Router, which can then route appropriate runtime states to it without introducing a new end-to-end workflow or modifying existing planners. Similarly, a new interaction mechanism can be added as an L3 Designer skill and registered in \texttt{widgets.json}. Once registered, existing L2 Planners can reuse the new interaction capability when its interaction goals, artifact constraints, return schema, and runtime requirements are compatible.

This organization also supports composition across task capabilities. Multiple task-specific L2 Planners can coexist in the same registry and operate over a shared runtime state containing user input, artifact state, interaction history, prior returns, and execution outcomes. As the task evolves, the L1 Router can route different states to different eligible planners within the same plugin. Meanwhile, L3 interaction capabilities and the shared L4 Renderer remain reusable across these planners. Thus, extending a plugin primarily adds or registers localized layer instances, while shared orchestration and interaction capabilities can continue to be composed across tasks without duplication.



%% file: sections/6-tech-evaluation.tex
\section{Technical Evaluation}\label{sec:tech_evaluation}
We further evaluate Inflak as a plugin across different host agent models to examine the stability and latency of executing its orchestration process during inference.
This section reports the experimental setup and results.

\subsection{Experiment Settings}

\paragraph{Dataset.} We constructed a dataset using the two Inflak plugins in Secs.~\ref{sec: gallery case 1} and Sec.~\ref{sec: gallery case 2}.
We tested the two plugins as (1) they include various tasks ranging from writing to image editing; (2) they are composed of multiple different interaction paradigms; and (3) they work in both chat-based and canvas-based interfaces.
For each plugin, we tested four resolution pathways mentioned in Sec.~\ref{sec:runtime pathways}, with five cases per pass type.
The coverage allows us to examine the stability and the latency comprehensively.
\revised{We used GPT-5.6 Sol~\cite{openai2026gpt56} to generate cases grounded in the plugin workflows with explicit preconditions. Each case specifies an input prompt, any required assets (e.g., a reference image for a poster), a simulated user response when interaction is needed (e.g., selecting a balanced stance), and the expected observable boundary with its evaluation criteria.
For example, one case pairs a fixed cafe photograph containing a bicycle and a chalkboard with a prompt asking the agent to let the user identify which object to remove before editing. The simulated response selects the chalkboard using a predefined bounding box; the evaluation records whether a selection interface is rendered and whether control subsequently returns to L1 router after submission.
Each case was repeated ten times with the same prompt, assets, and user response to assess run-to-run stability.}

\paragraph{Models and Metrics}
All cases are tested on two general-purpose LLMs within the GitHub Copilot agent, GPT-5.5~\cite{openai2026gpt55} and GPT-5.6 Sol~\cite{openai2026gpt56}, to examine whether Inflak can be executed consistently across different host agent models.
The stability of Inflak architecture is measured by \textit{layer compliance}, indicating whether each pass followed the layer order prescribed by the architecture.
For latency, we measured both \textit{per-layer} and \textit{end-to-end} values when executing resolution pathways with different models.
Intentionally, we did not measure the quality of finally generated interaction widgets or execution results, as they depend on the quality of the detailed widget design instructions in skills and the execution methods rather than Inflak orchestration alone.
Note that when computing values, we removed some runs that are not evaluable, such as those runs where agents did not invoke an Inflak plugin. 


\subsection{Results}

Overall, the results suggest that, as an agent extension, Inflak operates with manageable runtime latency and stable execution reliability across the two host agent models. We report pathway-level compliance and end-to-end latency in Table~\ref{tab:pathway_results}, and further decompose runtime cost by architectural layer in Table~\ref{tab:layer_latency}.

Regarding layer compliance, Host-Agent Return, Completion, and the
Execution Pathway achieved 100\% compliance for both models, as shown
in Table~\ref{tab:pathway_results}.
The Interaction Pathway achieved 94\% compliance for GPT-5.5 and 100\% for GPT-5.6 Sol.
The six noncompliant GPT-5.5 runs all occurred in visual collage authoring cases, with five skipping the L2 by transitioning directly from L1 to L3, while one transitioned backward from L3 to L2.
Overall, these results indicate that Inflak's prescribed layer progression can be followed reliably across the tested GPT host agents, including multi-layer execution and interaction pathways.

Among all pathways, Host-Agent Return and Completion had the shortest median end-to-end latencies of 3.06-4.39 seconds across the two models, as they only require running L1.
In comparison, the Interaction Pathway required median latencies of
30.12-30.87 seconds, while the Execution Pathway required
48.78-50.75 seconds.
The per-layer results in Table~\ref{tab:layer_latency} explain the observation.
Most layers other than the implementation one (L4) remained lightweight.
Across the two models, median latency was 3.86-3.92 seconds for L1, 6.58-7.08 seconds for L2, and 3.87-4.33 seconds for L3, showing an overall manageable time cost for the first three layers.
L4 was comparatively more expensive, with median latency ranging from 15.74 to 16.21 seconds.
This concentration of latency at L4 suggests that the additional cost of interaction-oriented pathways primarily arises during concrete interaction implementation rather than being uniformly distributed across the orchestration pipeline.
Similarly, execution pathways' latency mostly depends on the additional tools or model calls to execute tasks.


\begin{table*}[t]
\centering
\caption{
Pathway-level layer compliance and latency across host agent models under runtime constraints.
Latency reports the P95 (95th percentile) and median duration in seconds of resolving different resolution segments.
}
\label{tab:pathway_results}

\small
\setlength{\tabcolsep}{7pt}
\renewcommand{\arraystretch}{1.08}

\begin{tabular}{llcc}
\toprule
Model & Pathway
& Layer Compliance
& Latency: P95 / Median \\
\midrule

\multirow{4}{*}{GPT-5.5}
& Interaction
& 94/100 (94.00\%)
& 63.27 / 30.12 \\
& Execution
& 100/100 (100.00\%)
& 84.20 / 48.78 \\
& Host Agent Return
& 99/99 (100.00\%)
& 7.33 / 4.39 \\
& Completion
& 100/100 (100.00\%)
& 6.33 / 3.17 \\

\midrule

\multirow{4}{*}{GPT-5.6 Sol}
& Interaction
& 100/100 (100.00\%)
& 49.70 / 30.87 \\
& Execution
& 100/100 (100.00\%)
& 80.50 / 50.75 \\
& Host Agent Return
& 100/100 (100.00\%)
& 6.89 / 4.11 \\
& Completion
& 100/100 (100.00\%)
& 5.95 / 3.06 \\

\bottomrule
\end{tabular}
\end{table*}

\begin{table*}[t]
\centering
\caption{
Aggregated per-layer latency across host agent models under runtime constraints.
Each cell reports P95 (95th percentile) / median latency in seconds.
Note that L1 is executed across all four pathways. L2 is processed in the Interaction and Execution pathways, and L3-L4 are only invoked in the Interaction pathway.
}
\label{tab:layer_latency}

\small
\setlength{\tabcolsep}{7pt}
\renewcommand{\arraystretch}{1.08}

\begin{tabular}{lcccc}
\toprule
Model
& L1 Router
& L2 Planner
& L3 Designer
& L4 Renderer \\
\midrule

GPT-5.5
& 8.32 / 3.86
& 17.19 / 6.58
& 10.32 / 3.87
& 36.03 / 16.21 \\

GPT-5.6 Sol
& 10.26 / 3.92
& 27.49 / 7.08
& 8.70 / 4.33
& 25.84 / 15.74 \\

\bottomrule
\end{tabular}
\end{table*}

%% file: sections/7-case-study.tex
\section{Implementation Case Study}
\label{sec:implementation-case-study}

To examine how Inflak supports developers and interaction designers in implementing and evolving human-agent interaction, we conduct a comparative implementation case study on an existing interaction flow.
We compare two implementations of the same skill-oriented application: an Inflak-based plugin (Sec.~\ref{sec:instantiation}) and a vibe-coded plugin (VC) that similarly organizes task knowledge and interaction behavior through agent-readable skills and reusable resources.
We use the vibe-coded plugin as a comparison to reflect how a developer may implement the same functionality without following Inflak's layered organization.
\revised{Rather than comparing the finally produced interface quality or usability by two implementations, we focus on where the same interaction requirements are implemented, how responsibilities are distributed, and how changes propagate across components.}

\subsection{Case Scenario and Study Setup}

\paragraph{Scenario and requirements.}
Our case builds on the writing-revision and image-editing workflows in the existing application.
Both workflows can produce concrete artifact states or alternative outputs, but the original implementation does not provide a dedicated interaction for reviewing complete candidates side by side before deciding which one to retain.
We therefore extend the system with a \textit{Compare Viewer}, which presents multiple completed artifact states, preserves their identities, and allows users to select a preferred result with optional feedback.
\revised{We guided both systems with natural-language prompts addressing equivalent functional requirements, such as ``add a viewer for comparing and selecting completed candidates'' and, during iterative testing, ``fix the viewer not restoring the saved selection after reopening.''}
We formulate two requirements.
\textbf{R1: Adding a new interaction capability} introduces the Compare Viewer as a reusable interaction for reviewing and selecting among completed artifact states.
\textbf{R2: Integrating the capability into task workflows} allows the writing-revision and image-editing workflows to invoke the Compare Viewer when multiple completed candidates, or an original and edited result, should be compared before proceeding.
Together, they represent two common forms of interaction-flow evolution: introducing a new interaction capability and incorporating it into existing task workflows.

\paragraph{Compared systems and analysis method.}
We implement the above requirements in the Inflak-based plugin and the vibe-coded plugin.
\revised{We constructed the baseline vibe-coded plugin independently from a minimal GitHub Copilot SDK chat application, which is also used by the Inflak-based plugin.
Following a conventional vibe-coding workflow, we asked the agent to implement the same user-facing multimodal co-creation capabilities as the Inflak-based plugin.}
Both systems are modular and skill-based: Inflak realizes the four-layer orchestration architecture through extensible plugins, while the vibe-coded plugin organizes interaction behavior through task-oriented skills and shared resources.
To compare how the two systems accommodate capability extension and task integration, we distinguish between \textit{primary} (P) and \textit{dependent} (D) changes.
Primary changes directly define the requested interaction capability (R1) or the policy for invoking it in a task workflow (R2), whereas dependent changes adapt supporting or consuming components so that the new capability can be exposed and used.

\subsection{Comparative Change Analysis}

Figure~\ref{fig: implementation case} summarizes the file-level changes in the two implementations and distinguishes their primary and dependent roles.
In Inflak, the change is concentrated in one newly introduced interaction unit, one registry entry, and the task planners that consume it, without modifying any existing L3 interaction.
In VC, the primary implementation is largely concentrated in the shared \texttt{compare-viewer.md}, while integrating this specification requires dependent adaptations across the shared interaction skill and multiple task-specific skills.
Both implementations remain modular, but differ in how the new interaction capability is localized and how its integration propagates across the existing structure. We next unpack this contrast through the two requirements.

\subsubsection{R1. Adding a New Interaction Capability}

In Inflak, R1 is fulfilled with a new L3 Compare Viewer skill.
The skill defines the interaction for presenting completed artifact states, collecting a preferred candidate, and returning optional feedback.
The widget registry is dependently updated to register this new capability and make it available to compatible planners.
The implementation of the new interaction is therefore added as an independent unit rather than being incorporated into or redefining an existing interaction capability.
In VC, R1 is primarily expressed in the shared \texttt{compare-viewer.md}, which defines the comparison interaction, its expected inputs and returned information, and the corresponding interaction protocol together.
The existing \texttt{web-interaction} skill is then dependently updated to read this specification and realize the Compare Viewer during interaction generation.

\subsubsection{R2. Integrating the Capability into Task Workflows}

After R1, the Compare Viewer is available as an interaction capability.
R2 concerns when existing task workflows should invoke it.
In Inflak, this decision is owned by the two relevant L2 planners.
The writing-revision planner routes to the Compare Viewer when multiple complete revision candidates should be inspected before one is selected.
Similarly, the image-editing planner invokes it when multiple completed edits, or an original and edited artifact, should be compared.
The widget registry records the capability and its planner eligibility, allowing these planners to route to the new L3 interaction without changing its implementation or any other L3 skills.
In VC, the invocation policy is defined together with the interaction specification in \texttt{compare-viewer.md}.
The resulting dependency path is less uniform.
The shared \texttt{web-interaction/SKILL.md} reads the Compare Viewer reference to determine and generate the interaction.
At the same time, \texttt{writing-revision/SKILL.md} and \texttt{image-edit/SKILL.md} also directly reference the same Compare Viewer protocol (\texttt{web-interaction/references/compare-viewer.md}) while still using \texttt{web-interaction} at the same time.
Consequently, the task workflows depend on the capability through both a shared interaction skill and the underlying specification itself.
These paths are individually explicit, but the same capability is reached through different levels of abstraction, making its dependency structure less consistent.


\begin{figure}[htbp]
  \centering 
  \includegraphics[width=\columnwidth]{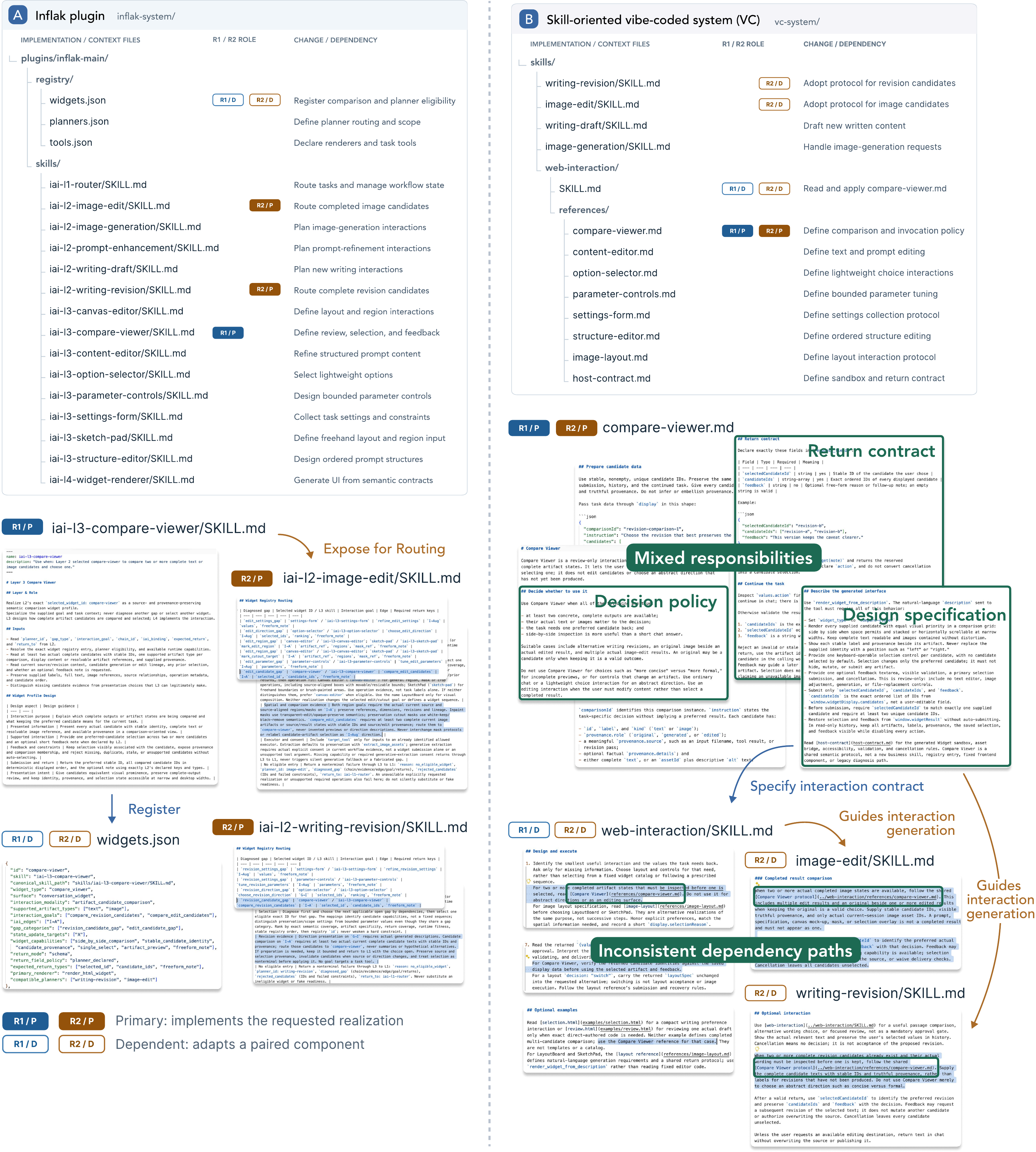}
  \caption{Comparison of implementation changes for adding artifact comparison to Inflak and the skill-oriented vibe-coded system (VC). 
  \revised{The upper panels show the implementation file structures and distinguish primary (P) from dependent (D) modifications under requirements R1 and R2. Below the horizontal division, the Inflak side (A) illustrates how the Compare Viewer is implemented, registered, and exposed to task-specific planners through modular components. In contrast, the VC side (B) shows how comparison logic, interface specifications, and interaction contracts are distributed across multiple files, resulting in mixed responsibilities and inconsistent dependency paths.}}
  \label{fig: implementation case}
  \vspace{-0.45cm}
\end{figure}

\subsection{Findings}

The case highlights three differences in how the two skill-based systems accommodate interaction-flow evolution.

\paragraph{Clear and stable responsibilities make the implementation easier to inspect.}
In Inflak, the mapping from the two requirements to implementation files is relatively direct.
R1 corresponds to one newly introduced L3 skill that owns the Compare Viewer realization, while R2 is handled by the two L2 planners that decide when the capability should be invoked.
The registry only exposes the capability and its eligibility.
This separation makes the implementation relatively intuitive to inspect.
Interaction design, task-specific invocation, and capability registration each have a focused location, and adding the new capability does not require modifying any existing interaction implementation.
VC also retains a modular skill-oriented file structure, but the extension introduces mixed responsibilities within \texttt{compare-viewer.md}.
As shown in Fig.~\ref{fig: implementation case}, \texttt{compare-viewer.md} contains the interaction design, the timing to use the interaction, and the return contract.
As a result, developers need to identify and understand intertwined functions from one single file.

\paragraph{Directional dependencies make change propagation easier to trace and contain.}
The two implementations also differ in how a newly introduced capability is linked to others.
In Inflak, the change follows a relatively explicit path: the new interaction capability is defined at L3, registered as an available capability, and consumed by the relevant L2 planners.
This provides a direct trace from introducing the capability to the task workflows affected by it.
In VC, the Compare Viewer is reached through two overlapping paths.
The task skills use \texttt{web-interaction} for interaction generation, while also directly reading the underlying \texttt{compare-viewer.md} protocol.
A change to the shared specification may therefore affect components that consume it directly as well as components that interpret it through \texttt{web-interaction}.
These dependencies remain traceable, but determining the propagation of a change requires understanding which parts of the specification are consumed through each path rather than following a single dependency chain.

\paragraph{Modularity with stable responsibility boundary and clear dependency definition facilitate later maintenance and reuse.}
The two features introduced above not only benefit developers' understanding but also make future change or reuse easier.
Both systems begin with modular, skill-oriented file structures, but the case illustrates how those boundaries evolve when a new cross-task interaction is introduced.
In Inflak, the Compare Viewer remains an independent L3 capability, while existing L3 interactions remain unchanged. 
Subsequently, new planners can reuse it through registration and routing without modifying its implementation.
Its interaction logic can therefore evolve separately from the task-specific conditions under which it is invoked.
In VC, reuse is also possible, but the mixed responsibility of \texttt{compare-viewer.md} and the coexistence of direct and mediated dependencies create additional coordination points.
For example, introducing the Compare Viewer into another workflow may require deciding whether the new task should directly adopt the shared protocol, rely on \texttt{web-interaction}, or both.
Updating the criteria to invoke the skill may also modify a file that defines interaction design and return behavior.
It therefore suggests that explicit responsibility boundaries and well-defined dependencies can help keep a single module more focused and independent for subsequent modification and reuse.

%% file: sections/8-discussion.tex
\section{Discussion}
We further reflect on the design choices and practical use of Inflak, and discuss opportunities for extending its orchestration capabilities in future human-agent systems.

\subsection{Reflection on Designing Inflak}
The key idea behind Inflak is to treat orchestrating interaction as a progressive resolution process with four layers.
This design \textit{facilitates the dynamic and in-situ characteristics of human-agent interaction} through disentangling multiple factors to consider for initiating interactions.
As illustrated in Sec.~\ref{sec:operationalize_iai}, based on a well-designed interaction model, we decompose the orchestration problem into three sub-problems: collecting contextual information to decide the current task, deciding whether to interact and what the interaction target is, and designing detailed interaction forms.
Through decomposing them, agents can have greater flexibility to decide what the final interaction would be, without prematurely binding to a specific interaction format at the design time.
For example, the multimodal co-creation case reuses a structured-form widget across writing and visual-design tasks, while combining it with task-specific interaction mechanisms (Sec.~\ref{sec:implementation-case-study}).
The flexibility is hard to be achieved by specifying all combinations following an end-to-end manner, as the number of combinations can reach the multiplication of agent tasks, interaction needs, and specific interaction formats.
Another benefit led by the progressive resolution is to \textit{lower the burden for developers to maintain}.
In our structure, agents are scaffolded to make decisions step-by-step. As a result, it is easier to monitor and diagnose which layer can be problematic.
And the fix can be applied only to a specific instance of the layer, without touching other instances.
The diagnosis and fix can be more challenging when an end-to-end interaction is triggered.

After deciding the usage of progressive orchestration approach with a layered structure, it is crucial to decide how many layers and what layers are responsible for.
Our final four-layer structure reflects a trade-off between efficiency and composability. 
Fewer layers can simplify the structure and reduce the time to finish one pathway, but also make it difficult to distinguish where routing, task reasoning, interaction design, and realization decisions originate, leading to low composability.
Conversely, decomposing orchestration into many small stages would make individual responsibilities precise at the cost of greater coordination overhead. 
Based on these considerations, we finally decided to adopt the four-layer structure as a balance point.
It allows us to separate collecting context information and deciding tasks (L1 Router), identifying interaction targets and goals (L2 Planner), making detailed interaction design (L3 Designer), and interaction widget implementation (L4 Renderer) into composable instances in layers.
L1 and L2 distinguish responsibility of information gathering, task decision, and interaction need judgment.
If merging L1 and L2, adding a new task would require re-implementing the entire L1's information collection function.
Furthermore, task selection will be left to the main agent without clear guidance.
L2 and L3 separate interaction goals and targets from the detailed formats.
If merging L2 and L3, the entanglement will lead to less flexibility to re-use interaction formats for different tasks.
L3 and L4 provide a clear boundary between design and implementation for maintainability.
If merging L3 and L4, transferring one Inflak plugin to another environment, such as switching from pure JavaScript-based to Vue-based web applications, would require significant changes in all L3 instances.
Our gallery in Sec.~\ref{sec: gallery} verifies that the four-layer structure is generally applicable and facilitates composing different interactions together.
Our technical evaluation (Sec.~\ref{sec:tech_evaluation}) shows that our layers can be processed at a reasonable time cost, which can be reduced through further optimization (see more discussion in Sec.~\ref{sec:future_work}).

\subsection{Usage of Inflak}
We envision Inflak primarily for agent developers and interaction designers who need to explicitly organize human participation within agent systems. 
It can be used when orchestrating human-agent interaction for a new task, such as the case in Sec.~\ref{sec:authoring_plugin}.
It is also possible to extend an existing agent with richer and more context-sensitive interactive capabilities (see Sec.~\ref{sec:implementation-case-study}).
We also envision that it provides a general solution to restructuring an existing agent workflow whose interaction logic has gradually become distributed across prompts, skills, and interface components for easier modification and extension.
Our examples in Sec.~\ref{sec: gallery} demonstrate this potential. 
We implemented interaction paradigms, which were originally created by different author teams, with Inflak universally.
Other than developers, Inflak can possibly serve as a common infrastructure for studying and testing new interaction design between humans and agents, as it requires little programming knowledge and low cost to update designs iteratively.
Furthermore, end users can use Inflak to achieve transferable interaction orchestration between different agents for the same interactive experience. 

Our current implementation implements Inflak as a skill-based agent extension. One benefit of this representation is that layer ownership does not impose a fixed granularity on individual instances. An L2 skill may cover a complete task such as image generation, or be decomposed into narrower planners for layout, style, or other concerns. Similarly, an L3 skill may represent an atomic interaction or a richer composition of related interaction mechanisms. These choices shift where orchestration complexity is managed. Broader instances keep more related knowledge and coordination locally within one skill, whereas finer instances expose capabilities more independently for modification and reuse, while requiring more coordination across skills. Natural-language skills accommodate both forms because they can flexibly package different amounts of task knowledge, policies, and constraints. Their self-contained form also allows Inflak to be packaged as an agent extension without modifying the host agent's core logic, suggesting opportunities for relatively lightweight reuse across compatible agent systems.

Beyond the granularity of each layer, agents' autonomy of implementing interaction designs can also be controlled when using Inflak.
Our current implementation primarily generates UI with L4 Renderer based on natural-language interaction specifications and structured contracts produced by L3 Designer.
We aim to preserve some visual design choices to be made by agents to enhance flexibility, leveraging their UI generation capability~\cite{cao2025generative, hojo2025generativeUI, leviathan2026generative}.
However, this is not the only choice provided by Inflak.
When widget consistency and reliability are prioritized, a predefined HTML/JavaScript/CSS template can be defined with L3 to fill in and L4 to render.
A descriptive specification, like most of our L3 instances (see Fig.~\ref{fig: plugin file structure} for example), can leave visual and interaction details to runtime generation.
Minimal specification with no specific design constraints can give the agent broader freedom for unforeseen interaction needs. 
This suggests that our method can be adapted to suit different scenarios with adjusting the autonomy level of agents.

Finally, other than skill-based plugins, layer responsibilities could also be implemented through other methods. 
For example, each layer could be assigned to a specialized agent, with information passed between these agents through explicit handoffs~\cite{qian2024chatdev}. 
In this design, the modular unit would shift from an individual skill specification to an independent agent responsible for a particular layer. 
Each agent could maintain its own context and reasoning process to take the corresponding responsibility.

\subsection{Future Work}\label{sec:future_work}
Inflak currently resolves orchestration primarily around the next interaction or execution requirement. More complex tasks may expose several gaps simultaneously or require multiple steps to be anticipated before execution can proceed. Supporting such cases would require L2 Planner to reason not only about the next gap, but also about the dependencies among multiple gaps and the sub-workflows needed to resolve them. This in turn raises a corresponding interaction-design problem at L3: multiple interaction requirements may need to be presented sequentially, in parallel, or through nested or composite widgets within a single generation. Different compositions can substantially change what users see at once, the order in which they provide information, and how one interaction constrains or reveals the next. Extending Inflak toward such multi-step orchestration therefore requires jointly reasoning about workflow structure and the composition of the interactions through which that workflow is exposed to users.

As orchestration accumulates across more tasks and interaction episodes, the resulting traces may also become a source of reusable orchestration knowledge. Inflak currently carries prior interaction and execution history forward as part of the runtime context, but does not explicitly abstract recurring patterns from this experience. Such traces could be used to derive lightweight user preferences when useful, but more importantly, they may reveal recurring relationships between task states, interaction needs, selected interaction mechanisms, and subsequent outcomes. Distilling these regularities into reusable knowledge could allow planners and interaction designers to draw on patterns learned from prior orchestration rather than relying entirely on manually authored policies. Over time, this may support a more evolving form of Inflak in which successful interaction patterns can be identified, generalized, and incorporated into future orchestration. This raises broader questions about how interaction traces should be abstracted into knowledge, how such knowledge should be represented and retrieved, and when newly distilled patterns should be incorporated into existing planning and interaction capabilities.

These extensions also make the efficiency of orchestration an important practical consideration. Our technical evaluation shows that latency varies across host models and resolution pathways, with L4 contributing a substantial portion of the overall runtime. Overall, however, the observed cost remains acceptable for the orchestration process demonstrated in our current implementation. Its relative impact may become even smaller in more complex agent workflows that already involve substantial model reasoning, retrieval, tool invocation, and repeated execution over longer periods. In such settings, the time spent resolving interaction orchestration can represent only a limited portion of the complete task process. Future work can further improve this efficiency by reducing repeated resolution, reusing previously established interaction structures, or executing Inflak asynchronously with a sub-agent.




%% file: sections/9-conclusion.tex
\section{Conclusion}
Dynamic and in-situ interaction facilitates smooth and productive collaboration between humans and AI agents.
This paper presents Inflak, a unified orchestration architecture to coordinate human-agent interaction with general applicability, runtime flexibility, module composability, and maintainability.
It applies a progressive resolution mechanism to decide when and how agents should interact with humans through instances of four different layers: router, planner, designer, and renderer.
We implement the Inflak architecture as skill-based plugins to agents since they are both agent-native and composable.
Inflak is evaluated through application galleries, technical measurement of latency and layer compliance, and a case study comparing adding new interactions to an Inflak-based plugin and a loosely structured plugin.
These evaluation results demonstrate Inflak can effectively meet the four requirements.
Looking ahead, Inflak can be further augmented by introducing multi-step planning, adapting to user preferences through self-evolving skills, and improving its runtime efficiency with caching and reusing.

%% file: sections/10-acknowledgement.tex
\begin{acks}
The authors wish to sincerely thank Ziao Liu, Leixian Shen, Yun Wang, Chang Han, and Yueyang Yu for their valuable feedback and insightful suggestions on this work.
\end{acks}